\documentclass[prb,reprint,preprintnumbers,amsmath,amssymb,showpacs,superscriptaddress,citeautoscript,aps]{revtex4-2}
\usepackage{amssymb}

\usepackage{graphicx}
\usepackage{float}
\usepackage{amsmath}
\usepackage{bm}
\usepackage[normalem]{ulem}
\usepackage{siunitx}
\usepackage{xcolor}
\usepackage{upgreek}
\usepackage{booktabs}
\usepackage{comment}

\usepackage{xr}
\makeatletter

\newcommand*{\addFileDependency}[1]{
\typeout{(#1)}
\@addtofilelist{#1}
\IfFileExists{#1}{}{\typeout{No file #1.}}
}\makeatother

\newcommand*{\myexternaldocument}[1]{%
\externaldocument[SI_]{#1}%
\addFileDependency{#1.tex}%
\addFileDependency{#1.aux}%
}

\myexternaldocument{SI}

\PassOptionsToPackage{hyphens}{url}
\usepackage{hyperref}  
\usepackage[all]{hypcap}

\begin{document}

\title{Nonequilibrium dynamics of high energy transitions in monolayer WSe${}_2$}

\author{Oleg Dogadov}
\affiliation{Department of Physics, Politecnico di Milano, Piazza Leonardo da Vinci 32, Milano, 20133, Italy}
\affiliation{Department of Physical Chemistry, Fritz Haber Institute of the Max Planck Society, 14195 Berlin, Germany}

\author{Jorge Cervantes-Villanueva}
\affiliation{Institute of Materials Science (ICMUV), University of Valencia, Catedr\'atico Beltrán 2, E-46980, Valencia, Spain}

\author{Nicholas Olsen}
\affiliation{Department of Chemistry, Columbia University, New York, 10027, NY, USA}

\author{Chiara Trovatello}
\affiliation{Department of Physics, Politecnico di Milano, Piazza Leonardo da Vinci 32, Milano, 20133, Italy}
\affiliation{Department of Mechanical Engineering, Columbia University, New York, 10027, NY, USA}

\author{Xiaoyang Zhu}
\affiliation{Department of Chemistry, Columbia University, New York, 10027, NY, USA}

\author{Giulio Cerullo}
\affiliation{Department of Physics, Politecnico di Milano, Piazza Leonardo da Vinci 32, Milano, 20133, Italy}
\affiliation{CNR-IFN, Piazza Leonardo da Vinci 32, Milan, 20133, Italy}

\author{Alejandro Molina-S\'anchez}
\affiliation{Institute of Materials Science (ICMUV), University of Valencia, Catedr\'atico Beltrán 2, E-46980, Valencia, Spain}

\author{Davide Sangalli}
\affiliation{Istituto di Struttura della Materia-CNR (ISM-CNR) and European Theoretical Spectroscopy Facility (ETSF), Piazza Leonardo da Vinci 32, 20133 Milano, Italy}

\author{Stefano Dal Conte}
\email{stefano.dalconte@polimi.it}
\affiliation{Department of Physics, Politecnico di Milano, Piazza Leonardo da Vinci 32, Milano, 20133, Italy}


\date{\today}

\begin{abstract}

High-energy optical transitions in monolayer transition-metal dichalcogenides exhibit characteristics that are markedly distinct from those of lower-lying band-edge excitons. 
These differences arise from the involvement of electronic states located at regions of the Brillouin zone that are displaced from the $K$ valleys. 
In this work, we investigate the ultrafast dynamics of these high-energy excitations by employing broadband ultrafast transient absorption spectroscopy spanning the visible to ultraviolet spectral range. 
We observe that the formation and relaxation dynamics of one of the high energy transitions display a distinct behavior compared to the lower-energy excitonic resonances, developing on a significantly slower timescale. 
First-principles calculations of the excitonic landscape allow us to account for this delayed response and attribute it to the phonon-mediated formation of momentum-dark excitons.

\end{abstract}

\maketitle


\section{Introduction}

Semiconducting transition metal dichalcogenides (TMDs) are layered materials that can be exfoliated down to a single layer (1L), transforming their band structure from an indirect to a direct bandgap. 
These materials are of particular interest in optoelectronics due to their strong light-matter interaction and their ability to absorb more than 10$\%$ of solar light at the 1L limit \cite{Palummo2015}. 
In these systems, strong quantum confinement along the out-of-plane direction and reduced Coulomb screening lead to the formation of strongly bound excitons, which are characterized by binding energies of the order of hundreds of meV \cite{Wang2018}.

Excitonic resonances dominate the optical response of 1L-TMDs, giving rise to absorption peaks that, in the limit of homogeneous broadening, have widths of only a few meV \cite{cadiz2017excitonic}. 
The lowest-energy excitonic peaks, which exhibit high oscillator strength, are commonly referred to as A and B excitons. 
They originate from electronic states at the $K$ and $K^\prime$ valleys, split by strong spin-orbit coupling, and can be selectively excited using circularly polarized light \cite{xiao2012coupled}. 
At higher energies in the visible (vis) and ultraviolet (UV) spectral range, additional resonances are observed \cite{wilson1969, beal1972transmission, zhao2013, li2014measurement, li2014broadband, song2019complex, ermolaev2020broadband, ermolaev2021spectroscopic}, which are associated with different optical transitions away from the $K$ points \cite{beal1972transmission, ramasubramaniam2012large, qiu2013optical, schmidt2016reversible, gu2019layer, woo2023excitonic}. 
At high energies, the presence of multiple transitions contributing to the same optical features makes it difficult to identify the dominant ones.

In recent years, considerable efforts have been devoted to the study of the non-equilibrium optical response of 1L-TMDs to better understand the recombination and relaxation processes of hot carriers and excitons. 
Optical pump-probe spectroscopy studies have revealed that the transient optical response of TMDs generally exhibits multi-component dynamics with timescales ranging from sub-picosecond for many-body effects and radiative carrier recombination \cite{aivazian2017manybody, trovatello2020ultrafast, genco2023ultrafast, genco2025ultrafast} to hundreds of picoseconds and even nanoseconds for dark- or defect states-assisted carrier recombination \cite{nie2019tailoring, morabito2024long}. 
The interplay of several concurrent processes, such as exciton formation \cite{steinleitner2017direct, trovatello2020ultrafast} and recombination \cite{robert2016exciton}, intra- and intervalley scattering \cite{dogadov2025dissecting}, exciton-exciton annihilation \cite{sun2014observation}, as well as many-body effects \cite{cunningham2017photoinduced, pogna2016photoinduced, trovatello2022disentangling, deckert2025coherent}, makes it particularly challenging to disentangle individual processes and determine their timescales.

\begin{figure*}[ht]
    \includegraphics[width = \textwidth]{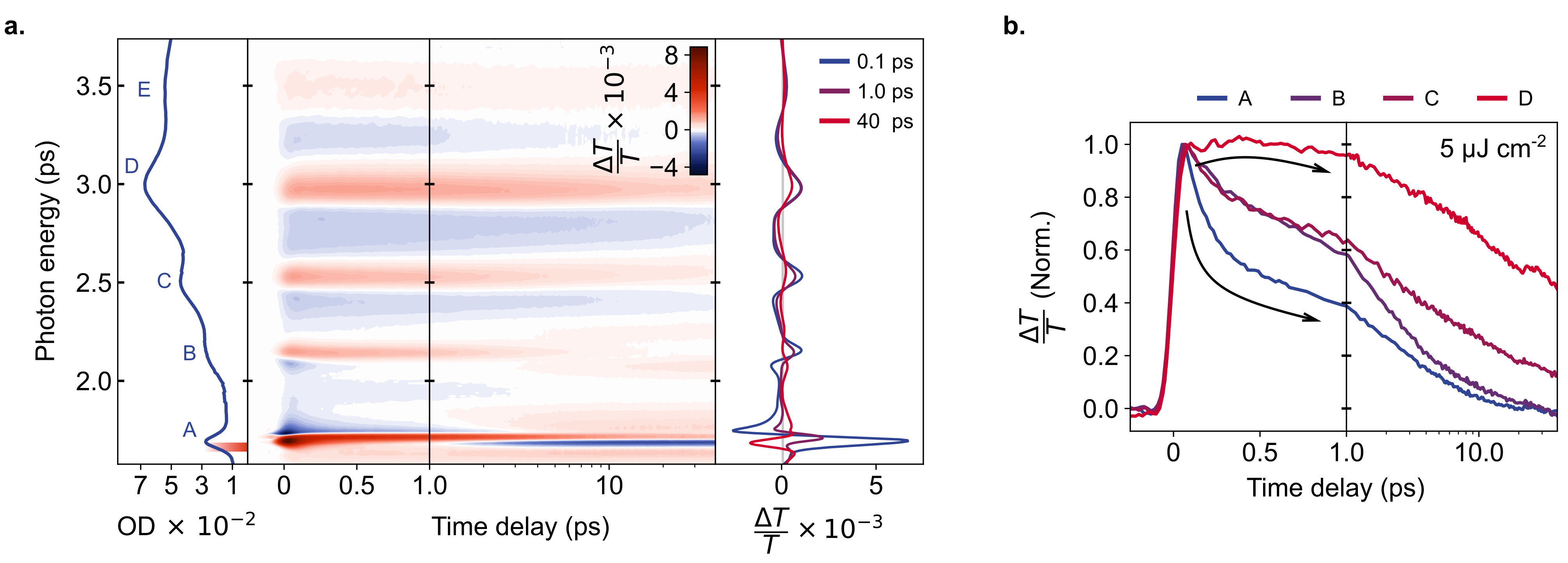}
    \caption{
    Transient optical response of 1L-$\mathrm{W Se_2}$ upon resonant pumping of A exciton (1.70~eV). 
    \textbf{a.} (Left) steady-state optical absorption spectrum in the vis-UV spectral range, showing several resonances; red line indicates the excitation energy.
    (Center) pseudo-color energy-time differential transmission $\Delta T/T$ map for 5.0~$\mathrm{\upmu J \, cm^{-2}}$ pump fluence. 
    (Right) $\Delta T/T$ spectra at select time delays. 
    \textbf{b.} Normalized time traces from map (a) showing dynamics around the maxima of A, B, C, and D transitions. 
    Black arrows are guides to the eye.
    The measurements are performed at 77~K.
    }
    \label{fig:1}
\end{figure*}

In the literature, most transient optical spectroscopy studies focus on the spectral region of the band-edge A and B excitons, with relatively few investigations targeting higher-energy resonances. 
Those that have, mainly on 1L-MoS${}_2$, show that high-energy resonances exhibit distinct relaxation dynamics compared to A and B excitons, due to different hot carrier relaxation pathways. 
A slower cooling dynamics for the C exciton, compared to band-edge excitons, was reported by Wang et~al. and attributed to the suppression of phonon-assisted recombination \cite{Wang2017}. 
Tran et~al. observed a delayed formation and slower relaxation dynamics of the D transition in 1L-MoS${}_2$, compared to other optical transitions, which was explained by its delayed generation mediated by intervalley-scattering up-conversion and cascade intervalley relaxation processes \cite{tran2024augmented}.
A similar slow dynamics of the D transition was also observed in few-layer WS${}_2$ \cite{Goswami2021} and organic super acid treated 1L-WSe${}_2$ \cite{chen2018investigating}, and attributed to slow intervalley scattering processes involving $K$ and $Q$ valleys \cite{Goswami2021}.
Despite the interest in long-lived states for optoelectronic applications \cite{tran2024augmented}, to date, a clear connection between the electronic states which are involved in the formation of the high energy optical transitions and their dynamics is still missing.        

In this work, we perform ultrafast transient absorption spectroscopy on 1L-WSe${}_2$ on an ultra-wide probe energy range, covering both the band-edge excitons and other transitions at higher energies. 
We find that, upon resonant excitation of the lowest energy bright transition, the dynamics of D excitonic transition around 3.0~eV displays an additional delayed build-up component compared to the instantaneous formation dynamics of the other excitonic transitions.
Based on our \textit{ab-initio} simulations of electronic and optical properties of 1L-WSe${}_2$, we explain this behavior by the intervalley scattering of photoexcited exciton population into energetically-favorable momentum-indirect dark excitons, which contribute to the observed delayed dynamics of the D transition. 


\section{Experimental results}

Experiments are performed on a large-area 1L-$\mathrm{WSe_2}$ on $\mathrm{SiO_2}$ substrate, fabricated with the gold-assisted mechanical exfoliation technique, as explained in Ref.~\cite{dogadov2025dissecting}.
Femtosecond transient absorption measurements are conducted in transmission geometry using an amplified Ti:sapphire laser at 2~kHz repetition rate.
For the resonant excitation of the sample, either the output of a tuneable home-built optical parametric amplifier or the second harmonic of the laser fundamental is used. 
In both cases, the $\sim \,$25~meV energy width of the pump pulses defines the $\sim \,$150~fs temporal resolution of our experiments.
At higher pump energies, the pulses become slightly longer as a result of increased second-order dispersion in the transmissive optical elements.
An ultra-broadband supercontinuum probe, generated in a $\mathrm{Ca F_2}$ plate and extending from 1.6 to 3.7~eV, tracks the dynamics of the A and B excitons and high energy states in 1L-$\mathrm{WSe_2}$. 
In all reported measurements, pump and probe beams are cross-linearly polarized and focused almost at normal incidence on the sample inside a cryostat.
The detailed information about the experimental setup is provided in the Supplementary Material (SM).

First, we measure the broadband transient response of 1L-$\mathrm{WSe_2}$ upon resonant excitation of A exciton 1$s$ state (1.70~eV) at 77~K. 
Figure~\ref{fig:1} summarizes the experimental results. 
The differential transmission $\Delta T/T$ map in Figure~\ref{fig:1}a, acquired for 5.0~$\mathrm{\upmu J \, cm^{-2}}$ pump fluence, corresponding to $\sim \! 5 \times 10^{11} \ \mathrm{cm^{-2}}$ photocarrier density, which is at least an order of magnitude lower than the expected Mott transition threshold \cite{chernikov2015population, steinhoff2017exciton}, reveals several positive signals in the probed range, appearing close to the resonances observed in the steady-state optical absorption spectrum, shown on the left.
Apart of the well-studied A and B excitons at 1.70~eV and 2.15~eV, respectively, originating from the spin-orbit split states at the corners of the Brillouin zone (BZ), three features at higher energy are observed, which we label for simplicity C ($\sim \! 2.5$~eV), D ($\sim \! 3.0$~eV), and E ($\sim \! 3.5$~eV) transitions.

As reported previously, the resonant pumping of A exciton leads to a rapid formation of the transient signal for the transitions at higher energies \cite{trovatello2022disentangling, pogna2016photoinduced}.
\begin{figure}[ht]
    \includegraphics[width = .48\textwidth]{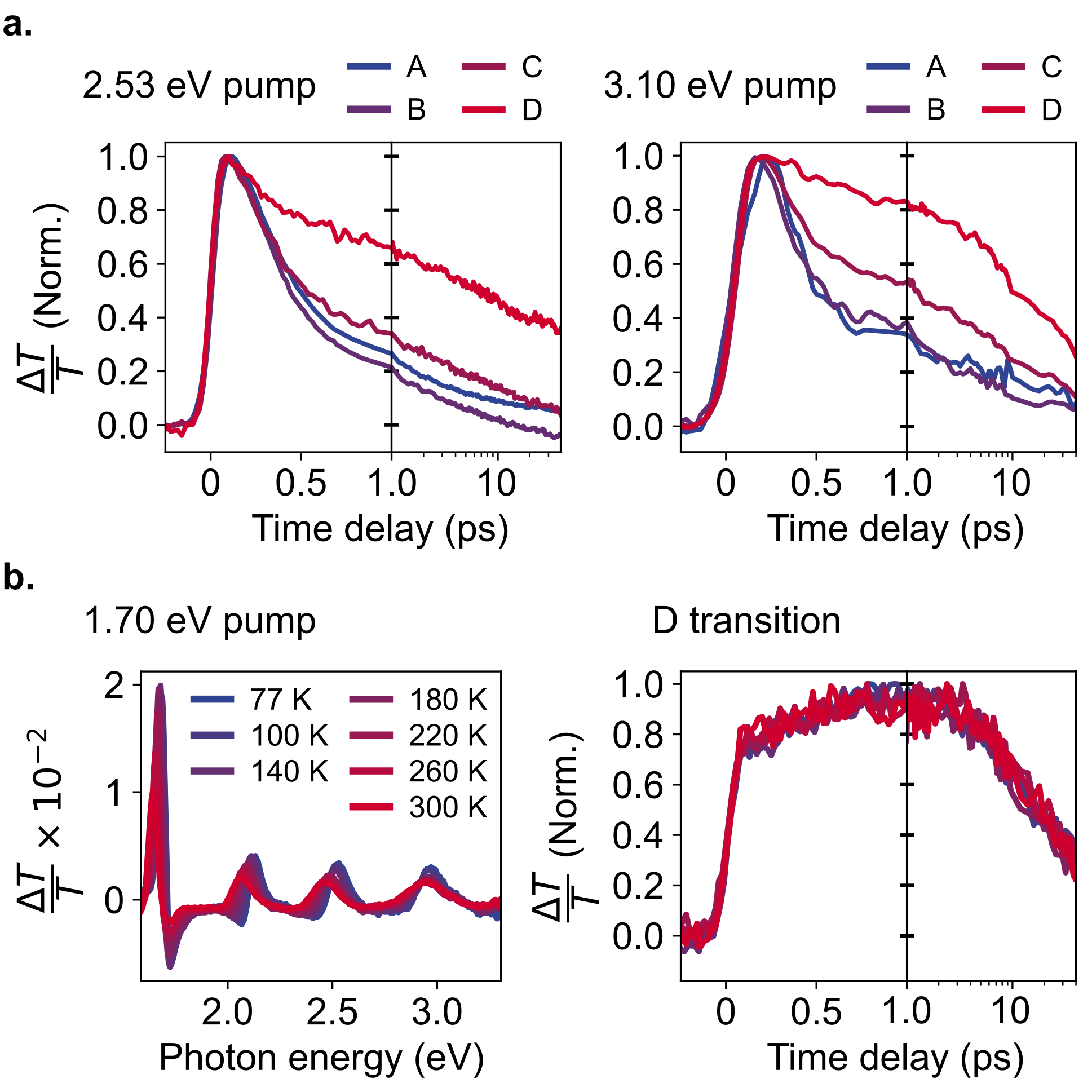}
    \caption{
    Pump energy and sample temperature effect. 
    \textbf{a.} Normalized time traces of A, B, C, and D transitions upon pumping C (2.53~eV, left) and D transition (3.10~eV, right) (cf. Figure~\ref{fig:1}b).
    Pump fluence 5.0~$\mathrm{\upmu J \, cm^{-2}}$ for both pump photon energies; the measurements are performed at 77~K.
    \textbf{b.} Temperature dependence. 
    (Left) transient spectra at 100~fs time delay and (right) D transition time traces.
    Resonant A exciton pump (1.70~eV), 15~$\mathrm{\upmu J \, cm^{-2}}$ pump fluence.
    }
    \label{fig:2}
\end{figure}
The dynamics of different signals, however, is not identical. 
The signals of A and B excitons have complex temporal evolution of the line shapes (see the right panel in Figure~\ref{fig:1}a) due to the sophisticated interplay of many-body effects and population dynamics \cite{trovatello2022disentangling, Calati2023, deckert2025coherent}, as well as thermal effects \cite{Calati2023}. 
In contrast, the high energy transitions show almost negligible evolution of the signal shapes in the explored time window, although some narrow-band transitions are expected in that energy region \cite{lin2021}.
Comparing the time traces at different probe photon energies shown Figure~\ref{fig:1}b, we find that in contrast to other traces the D transition has a delayed component:
An instantaneous rise, which is not resolved within our experimental temporal resolution, is followed by a delayed rise and a plateau, whereas the other transitions, following a pulsewidth-limited rise, undergo a multicomponent decay. 
A comparison with the E transition dynamics can be found in the SM \cite{supplementary_citations}. 

As shown in Figure~\ref{fig:2}a, when the 1L-$\mathrm{WSe_2}$ is excited at high energy (2.53~eV or 3.10~eV), the delayed rise of the D transition signal is absent, although its relaxation dynamics is still considerably longer than that of the other transitions. 
The delayed rise component, seen in the time trace in Figure~\ref{fig:1}b, is unique for the resonant pumping of A exciton, indicating a presence of an additional scattering channel between the photoexcited A excitons and the states contributing to the D transition.

We further explore the effect of sample temperature on the observed delayed dynamics.
As the temperature increases, all transitions progressively shift towards lower energies (Figure~\ref{fig:2}b left), as a result of the enhanced electron-phonon interaction \cite{odonnell1991, nguyen2024}.
Across the explored temperature range, the dynamics of the D transition remain essentially unchanged, as evidenced by the nearly perfect overlap of all time traces shown in Figure~\ref{fig:2}b (right).
The observed temperature independence of the dynamics implies that the formation of the signal is not mediated by the absorption of thermally-activated phonons, and other processes dominate the delayed formation of the signal. 

To identify possible channels governing the short time dynamics in the high energy range, we analyze the contributions of electronic states in 1L-$\mathrm{WSe_2}$ to the observed optical transitions.
For this, we perform a theoretical analysis of the 1L-$\mathrm{WSe_2}$ band structure, on which we focus in the following section. 


\section{Numerical simulations}

First-principles calculations are carried out within the framework of many-body perturbation theory (MBPT). Electronic properties considering quasiparticle corrections are obtained using the GW approximation, while optical properties taking into account excitonic effects are evaluated by solving the Bethe–Salpeter Equation (BSE) (detailed computational methods are provided in the SM, while the data supporting the findings are openly available \cite{dogadov_2026_19815236}).

\begin{figure}[b]
    \includegraphics[width = 0.48\textwidth]{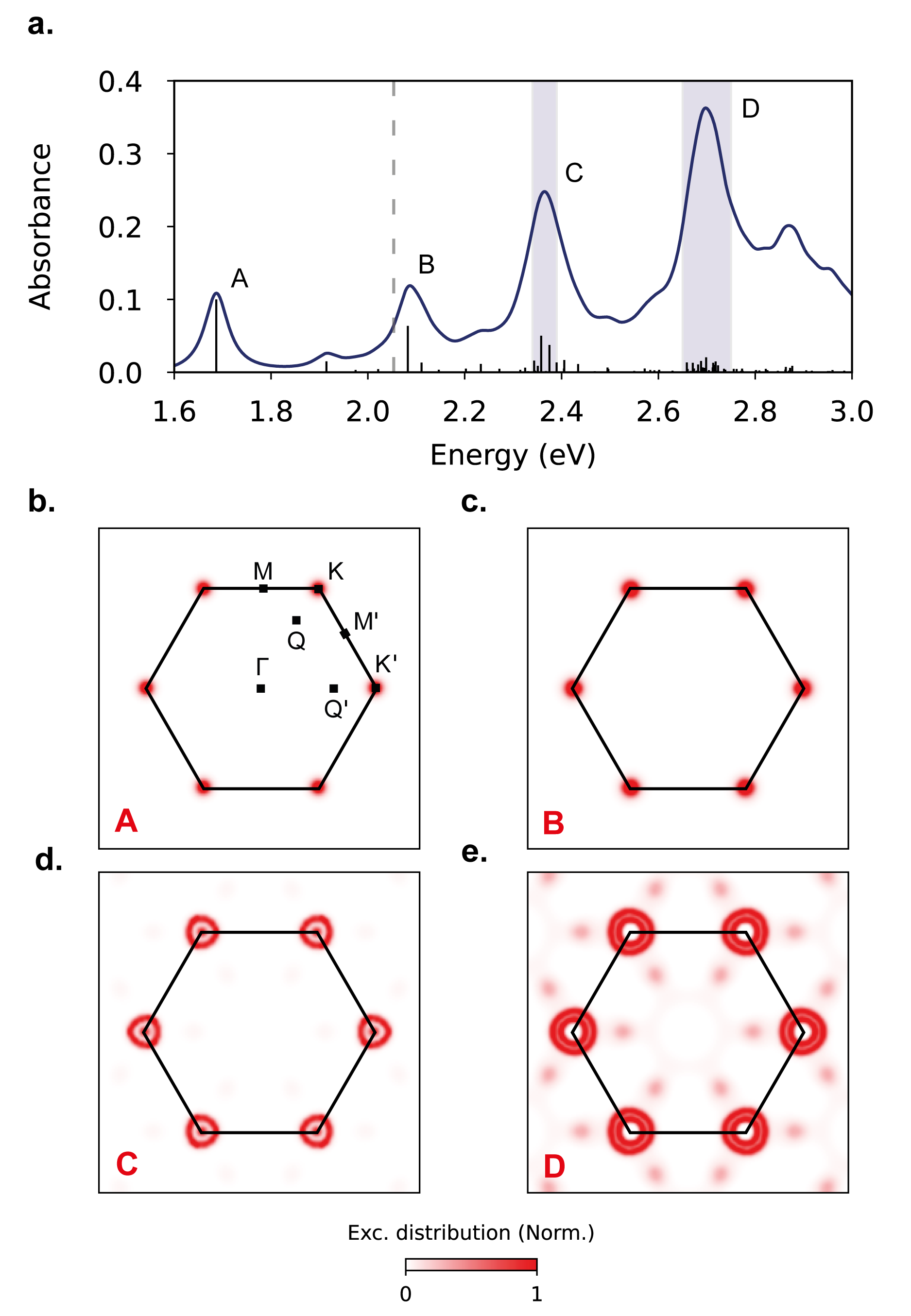}
    \caption{Theoretical analysis of excitons across the Brillouin zone (BZ). 
    (a) Calculated absorption spectrum of 1L-WSe${}_2$. 
    The vertical dashed line indicates the electronic bandgap obtained from GW calculations. 
    The vertical black lines represent the electronic transitions, with their heights proportional to the corresponding oscillator strengths. 
    Energy windows of $\pm 25$~meV and $\pm 50$~meV around the maxima of the C and D peaks, respectively, highlighted by the filled regions, are considered.
    (b-e) Distribution of the direct optical transitions contributing to individual peaks across the BZ: (b) A, (c) B, (d) C, and (e) D peaks. 
    Contributions for each peak are normalized.}
    \label{fig:3}
\end{figure}

Figure~\ref{fig:3}(a) shows the equilibrium absorption spectrum with relevant excitonic peaks labeled as A, B, C, and D, along with the corresponding oscillator strengths. 
The simulated spectrum shows overall similarity to the experimental one, reported in Figure~\ref{fig:1}a. 
The C and D peaks arise from a manifold of electronic transitions, in contrast to the A and B peaks, which are dominated by individual excitonic transitions. 
To take into account the contribution of multiple states to the optical transitions, energy windows of $\pm 25$~meV and $\pm 50$~meV around the C and D peaks, respectively, are considered.

To investigate the origin of the different formation times at different energies in the experimental transient signal, the excitonic wavefunctions of the states contributing to the absorption peaks are mapped throughout the BZ as depicted in Figures~\ref{fig:3}(b–e). 
The figure shows the distribution of direct optical transitions contributing to the corresponding features observed in the absorption spectrum in Figure~\ref{fig:3}a. 
The intensity indicates relative contribution of individual transitions. 

The A and B excitons, originating from the spin-orbit split bands at the corners of the BZ, are mainly localized at the $K$/$K^\prime$ valleys. 
In contrast, the transitions contributing to the C peak are mainly distributed around the $K$/$K^\prime$ valleys, forming a characteristic ring-like pattern, with an additional minor contribution near the $Q$/$Q^\prime$ valleys between $K$/$K^\prime$ and $\Gamma$ points of the BZ.
Similarly, the pattern for the D peak displays a well-defined double ring centered around the $K$/$K^\prime$ valley. 
However, the contributions of transitions around the $Q$/$Q^\prime$ valley is significantly higher than for other peaks.
These results are in agreement with the simulations reported in \cite{woo2023excitonic}.

\begin{figure}[t!]
    \begin{center}        
    \includegraphics[width = 0.475\textwidth]{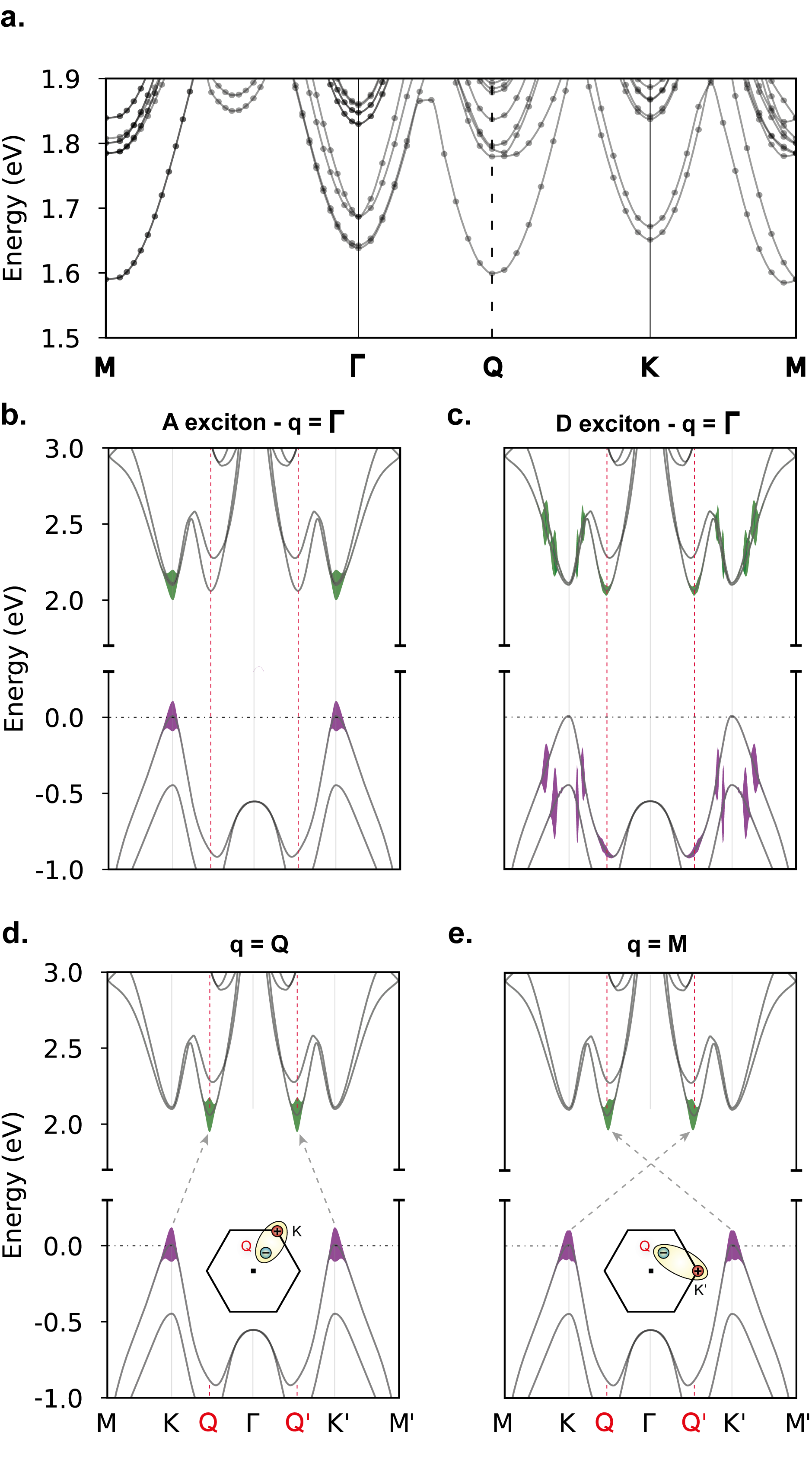}
    \end{center}
    \caption{Momentum-dependence of the excitons along the BZ. 
    (a) Exciton dispersion as a function of the momentum. The vertical dashed line represents center-of-mass momentum exciton with \textbf{q} = \textbf{Q}. 
    (b-e) Excitons along the band structure for (b) A exciton with \textbf{q} = $\mathbf{\Gamma}$, (c) D exciton with \textbf{q} = $\mathbf{\Gamma}$, (d) lowest energy exciton with \textbf{q} = \textbf{Q} and, (e) lowest energy exciton with \textbf{q} = \textbf{M}. 
    By symmetry, $\mathbf{q} = \mathbf{Q}'$ and $\mathbf{q} = \mathbf{M}'$ excitons have the same momentum dependence.    
    The purple areas represent the hole at the valence band maximum, while the green areas indicate the excited electron at the conduction band minimum. 
    The diagonal arrows represent the center-of-mass momentum while the vertical dashed red lines represent the $Q$ and $Q^\prime$ valleys.}
    \label{fig:4}
\end{figure}

In addition to optically active transitions with vanishing center-of-mass momentum, i.e., \textbf{q} = $\mathbf{\Gamma}$, which are directly accessible in absorption measurements, we also consider transitions with finite center-of-mass exciton momentum $\mathbf{q}$. 
(From this point onward, bold symbols will denote finite center-of-mass momenta of the exciton, whereas regular symbols will refer to the electron or hole momenta.) 
Excitons at finite center-of-mass momentum are optically dark; however, they can be populated via phonon-mediated relaxation pathways following optical excitation \cite{Malic2018, Selig2018}. 
The calculated exciton dispersion shown in Figure~\ref{fig:4}(a) exhibits multiple local minima, each corresponding to distinct finite-momentum excitonic transitions involving electron–hole pairs residing in different valleys of the BZ. 
To simplify the nomenclature, we refer to $\mathbf{q}=\mathbf{Q}$ and $\mathbf{q}=\mathbf{M}$ excitons in the following. 
The same discussion applies also to $\mathbf{q}=\mathbf{Q}'$ and $\mathbf{q}=\mathbf{M}'$ excitons. 
Once A exciton population is generated by resonant photoexcitation, it relaxes via phonon-mediated scattering towards the global minima of the exciton dispersion at $\mathbf{q}$ = $\mathbf{Q}$ (connecting the $K$ ($K^\prime$) and $Q$ ($Q^\prime$) valleys) and $\mathbf{q}$ = $\mathbf{M}$ (connecting the $K$ ($K^\prime$) and $Q^\prime$ ($Q$) valleys). 
In Figure~\ref{fig:4}(b-c) we show the wavefunction for the optically active ($\mathbf{q}=\pmb{\Gamma}$) A exciton and D peak projected on the electronic band structure, and compare them with the lowest energy excitons at $\mathbf{q}=\mathbf{Q,M}$, as shown in Figure~\ref{fig:4}(d-e).
By comparing the zero-momentum and finite-momentum excitons presented in Figure~4, we find that the zero momentum A exciton shares the same hole state in the $K$/$K^\prime$ valleys as the optically dark finite-momentum excitons at the $\mathbf{M}$ and $\mathbf{Q}$ points, whereas the zero momentum D transition shares the electronic state in the $Q$/$Q^\prime$ valleys. 
These distinct state overlaps account for the delayed formation dynamics observed in the pump–probe signal at the D peak.


\section{Discussion and conclusion}

Under resonant excitation, the A exciton is immediately bleached due to Pauli blocking and many-body effects that lead to an energy renormalization of the excitonic transition. 
Efficient phonon-mediated scattering of electrons from the $K$/$K^\prime$ valleys in the conduction band to the $Q$/$Q^\prime$ valleys, at slightly lower energy, partially converts the bright exciton population into finite-momentum dark excitons. 
The electronic component of these excitons induces Pauli blocking of the D excitonic transition, giving rise to their delayed bleaching signal.
The timescale of this signal is in excellent agreement with that of the formation of momentum-dark excitons, measured directly via time- and angle-resolved photoemission spectroscopy \cite{Madeo2020, Bange2023}, thereby confirming the transient origin of the signal associated with the D transition.

The lack of temperature dependence of the signal build-up timescale is attributable to the process governing the electronic population transfer between the valleys, which is mediated by spontaneous phonon emission.
Given the small energy separation between the $K$/$K^\prime$ and $Q$/$Q^\prime$ valleys, the scattering process is dominated by low-energy phonon-assisted processes, which are sufficient to compensate for the energy mismatch between the electronic states involved.
On the other hand, the quasi-instantaneous component of the D transition build-up dynamics originates from an ultrafast renormalization of the electronic-state energies, driven by a prompt modification of the electron–hole interaction \cite{Sangalli2016}.

Our simulations indicate that the bandgap in 1L-WSe${}_2$ is indirect, as suggested for 1L-WSe${}_2$ on SiO${}_2$ substrate \cite{wang2024experimental}.
However, the debate regarding whether this material possesses a direct or indirect bandgap (subject to such factors as the dielectric environment, strain, or sample temperature) persists in the literature \cite{tonndorf2013photoluminescence, hsu2017evidence, erben2018excitation, muoi2019electronic, Madeo2020, wang2024experimental}.
We note, however, that our conclusions about the delayed dynamics of the D transition rely on the energetic proximity of the states in $K$/$K^\prime$ and $Q$/$Q^\prime$ valleys in the conduction band, as well as the efficiency of the intervalley scattering channel \cite{Madeo2020, dogadov2025dissecting}, rather than the absolute energy ordering of the global bandgap minimum.
Furthermore, upon photoexcitation, the $Q$/$Q^\prime$ valleys can undergo larger energy shifts than the $K$/$K^\prime$ ones \cite{erben2018excitation, steinhoff2015efficient}, making the intervalley scattering more favorable. 
 
In summary, we investigate the dynamics of the high-energy interband transitions in 1L-$\mathrm{WSe_2}$ and find that, upon resonant excitation at the optical gap, the D transition exhibits a distinctly delayed formation on sub-picosecond timescale, in contrast to the nearly instantaneous build-up of the lower-energy excitonic transitions. 
The theoretical analysis of the excitonic landscape across the BZ, including its momentum dependence, shows that this transition has negligible contributions from electronic states at the $K$/$K^\prime$ valleys and significant contributions from states at the $Q$/$Q^\prime$ valleys. 
It also shares the same conduction-band electronic states as the finite-momentum \textbf{q} = \textbf{Q}, \textbf{M} excitons. 
The close energetic proximity between the bright A excitons and these dark excitons facilitates their population and explains the observed delayed formation dynamics of the D transition.  
High-energy excitonic transitions in monolayer TMDs provide crucial insight into electronic states beyond the $K$ valleys, enabling direct access to momentum-indirect and dark excitons that govern ultrafast carrier relaxation and intervalley scattering. 
These states play a central role in defining the broadband optical response, hot-carrier dynamics, and many-body interactions that emerge on sub-picosecond timescales. Understanding the formation and coupling of high-energy excitons is therefore essential not only for building a complete picture of the excitonic landscape, but also for advancing technological applications that rely on efficient energy redistribution at high fields and photon energies. 


\begin{acknowledgments}

O.D. and G.C. acknowledge financial support from European Union's NextGenerationEU Investment 1.1, PRIN 2022 PNRR HAPPY [ID P20224AWLB, CUP D53D23016720001]. 
G.C. and S.D.C. acknowledge support from the European Union’s Next Generation EU Programme with the I-PHOQS Infrastructure 423 [IR0000016, ID D2B8D520, CUP B53C22001750006] “Integrated infrastructure initiative in Photonic and Quantum Sciences”. 
G.C., S.D.C., and C.T. acknowledge support by the acknowledge the Horizon Europe European Innovation Council (101130384, HORIZONEIC-2023-PATHFINDEROPEN-01, QUONDENSATE). 
S.D.C. acknowledges support from the European Union's NextGenerationEU – Investment 1.1, M4C2 - Project n. 2022LA3TJ8 – CUP D53D23002280006.
C.T. and G.C. acknowledge funding from the European Union—NextGenerationEU under the National Quantum Science and Technology Institute (NQSTI) Grant No. PE00000023-q-ANTHEM-CUP H43C22000870001. 
C.T. acknowledges the European Union’s Horizon Europe research and innovation programme under the Marie Sk{\l}odowska-Curie PIONEER HORIZON-MSCA-2021-PF-GF grant agreement No~101066108. 
D.S. acknowledges funding from the “MAterials design at the eXascale” (MaX) center of excellence, co-funded by the European High Performance Computing joint Undertaking (JU) and participating countries (Grant Agreement No. 101093374), and from the PRIN project “Exploring extreme ultra-violet excitons with attosecond time resolution” (EXATTO), Grant No. 2022PX279E from MIUR (Italy).
X.Y.Z acknowledges support from Programmable Quantum Materials, an Energy Frontier Research Center funded by the U.S. Department of Energy (DOE), Office of Science, Basic Energy Sciences (BES), under award DE-SC0019443.
This work is supported by the Horizon Europe research and innovation program of the European Union under the Marie Sk{\l}odowska-Curie grant agreement 101118915 (TIMES). 
This work is part of the project I+D+i PID2023-146181OB-I00 UTOPIA, funded by MCIN/AEI/10.13039/501100011033, the project PROMETEO/2024/4 (EXODOS) and SEJIGENT/2021/034 (2D-MAGNONICS) funded by the Generalitat Valenciana. 
This study is also part of the Advanced Materials program (project SPINO2D), supported by MCIN with funding from the European Union NextGenerationEU (PRTR-C17.I1) and Generalitat Valenciana. 
J.C.V and A.M.S. thankfully acknowledges the computer resources at Agustina and technical support provided by BIFI and Barcelona Supercomputing Center (FI-2025-2-0001), and the computer resources at Tirant-UV (project lv48 - FI-2025-2-0001).

\end{acknowledgments}

\section*{Data availability statement}

The data supporting the findings of this study are publicly available in the Zenodo repository at https://zenodo.org/records/19815236


\begin{thebibliography}{73}%
\makeatletter
\providecommand \@ifxundefined [1]{%
 \@ifx{#1\undefined}
}%
\providecommand \@ifnum [1]{%
 \ifnum #1\expandafter \@firstoftwo
 \else \expandafter \@secondoftwo
 \fi
}%
\providecommand \@ifx [1]{%
 \ifx #1\expandafter \@firstoftwo
 \else \expandafter \@secondoftwo
 \fi
}%
\providecommand \natexlab [1]{#1}%
\providecommand \enquote  [1]{``#1''}%
\providecommand \bibnamefont  [1]{#1}%
\providecommand \bibfnamefont [1]{#1}%
\providecommand \citenamefont [1]{#1}%
\providecommand \href@noop [0]{\@secondoftwo}%
\providecommand \href [0]{\begingroup \@sanitize@url \@href}%
\providecommand \@href[1]{\@@startlink{#1}\@@href}%
\providecommand \@@href[1]{\endgroup#1\@@endlink}%
\providecommand \@sanitize@url [0]{\catcode `\\12\catcode `\$12\catcode
  `\&12\catcode `\#12\catcode `\^12\catcode `\_12\catcode `\%12\relax}%
\providecommand \@@startlink[1]{}%
\providecommand \@@endlink[0]{}%
\providecommand \url  [0]{\begingroup\@sanitize@url \@url }%
\providecommand \@url [1]{\endgroup\@href {#1}{\urlprefix }}%
\providecommand \urlprefix  [0]{URL }%
\providecommand \Eprint [0]{\href }%
\providecommand \doibase [0]{https://doi.org/}%
\providecommand \selectlanguage [0]{\@gobble}%
\providecommand \bibinfo  [0]{\@secondoftwo}%
\providecommand \bibfield  [0]{\@secondoftwo}%
\providecommand \translation [1]{[#1]}%
\providecommand \BibitemOpen [0]{}%
\providecommand \bibitemStop [0]{}%
\providecommand \bibitemNoStop [0]{.\EOS\space}%
\providecommand \EOS [0]{\spacefactor3000\relax}%
\providecommand \BibitemShut  [1]{\csname bibitem#1\endcsname}%
\let\auto@bib@innerbib\@empty
\bibitem [{\citenamefont {Palummo}\ \emph {et~al.}(2015)\citenamefont
  {Palummo}, \citenamefont {Bernardi},\ and\ \citenamefont
  {Grossman}}]{Palummo2015}%
  \BibitemOpen
  \bibfield  {author} {\bibinfo {author} {\bibfnamefont {M.}~\bibnamefont
  {Palummo}}, \bibinfo {author} {\bibfnamefont {M.}~\bibnamefont {Bernardi}},\
  and\ \bibinfo {author} {\bibfnamefont {J.~C.}\ \bibnamefont {Grossman}},\
  }\bibfield  {title} {\bibinfo {title} {Exciton radiative lifetimes in
  two-dimensional transition metal dichalcogenides},\ }\href
  {https://doi.org/10.1021/nl503799t} {\bibfield  {journal} {\bibinfo
  {journal} {Nano Letters}\ }\textbf {\bibinfo {volume} {15}},\ \bibinfo
  {pages} {2794 – 2800} (\bibinfo {year} {2015})}\BibitemShut {NoStop}%
\bibitem [{\citenamefont {Wang}\ \emph {et~al.}(2018)\citenamefont {Wang},
  \citenamefont {Chernikov}, \citenamefont {Glazov}, \citenamefont {Heinz},
  \citenamefont {Marie}, \citenamefont {Amand},\ and\ \citenamefont
  {Urbaszek}}]{Wang2018}%
  \BibitemOpen
  \bibfield  {author} {\bibinfo {author} {\bibfnamefont {G.}~\bibnamefont
  {Wang}}, \bibinfo {author} {\bibfnamefont {A.}~\bibnamefont {Chernikov}},
  \bibinfo {author} {\bibfnamefont {M.~M.}\ \bibnamefont {Glazov}}, \bibinfo
  {author} {\bibfnamefont {T.~F.}\ \bibnamefont {Heinz}}, \bibinfo {author}
  {\bibfnamefont {X.}~\bibnamefont {Marie}}, \bibinfo {author} {\bibfnamefont
  {T.}~\bibnamefont {Amand}},\ and\ \bibinfo {author} {\bibfnamefont
  {B.}~\bibnamefont {Urbaszek}},\ }\bibfield  {title} {\bibinfo {title}
  {Colloquium: Excitons in atomically thin transition metal dichalcogenides},\
  }\bibfield  {journal} {\bibinfo  {journal} {Reviews of Modern Physics}\
  }\textbf {\bibinfo {volume} {90}},\ \href
  {https://doi.org/10.1103/RevModPhys.90.021001} {10.1103/RevModPhys.90.021001}
  (\bibinfo {year} {2018})\BibitemShut {NoStop}%
\bibitem [{\citenamefont {Cadiz}\ \emph {et~al.}(2017)\citenamefont {Cadiz},
  \citenamefont {Courtade}, \citenamefont {Robert}, \citenamefont {Wang},
  \citenamefont {Shen}, \citenamefont {Cai}, \citenamefont {Taniguchi},
  \citenamefont {Watanabe}, \citenamefont {Carrere}, \citenamefont {Lagarde},
  \citenamefont {Manca}, \citenamefont {Amand}, \citenamefont {Renucci},
  \citenamefont {Tongay}, \citenamefont {Marie},\ and\ \citenamefont
  {Urbaszek}}]{cadiz2017excitonic}%
  \BibitemOpen
  \bibfield  {author} {\bibinfo {author} {\bibfnamefont {F.}~\bibnamefont
  {Cadiz}}, \bibinfo {author} {\bibfnamefont {E.}~\bibnamefont {Courtade}},
  \bibinfo {author} {\bibfnamefont {C.}~\bibnamefont {Robert}}, \bibinfo
  {author} {\bibfnamefont {G.}~\bibnamefont {Wang}}, \bibinfo {author}
  {\bibfnamefont {Y.}~\bibnamefont {Shen}}, \bibinfo {author} {\bibfnamefont
  {H.}~\bibnamefont {Cai}}, \bibinfo {author} {\bibfnamefont {T.}~\bibnamefont
  {Taniguchi}}, \bibinfo {author} {\bibfnamefont {K.}~\bibnamefont {Watanabe}},
  \bibinfo {author} {\bibfnamefont {H.}~\bibnamefont {Carrere}}, \bibinfo
  {author} {\bibfnamefont {D.}~\bibnamefont {Lagarde}}, \bibinfo {author}
  {\bibfnamefont {M.}~\bibnamefont {Manca}}, \bibinfo {author} {\bibfnamefont
  {T.}~\bibnamefont {Amand}}, \bibinfo {author} {\bibfnamefont
  {P.}~\bibnamefont {Renucci}}, \bibinfo {author} {\bibfnamefont
  {S.}~\bibnamefont {Tongay}}, \bibinfo {author} {\bibfnamefont
  {X.}~\bibnamefont {Marie}},\ and\ \bibinfo {author} {\bibfnamefont
  {B.}~\bibnamefont {Urbaszek}},\ }\bibfield  {title} {\bibinfo {title}
  {{Excitonic Linewidth Approaching the Homogeneous Limit in
  ${\mathrm{MoS}}_{2}$-Based van der Waals Heterostructures}},\ }\href
  {https://doi.org/10.1103/PhysRevX.7.021026} {\bibfield  {journal} {\bibinfo
  {journal} {Phys. Rev. X}\ }\textbf {\bibinfo {volume} {7}},\ \bibinfo {pages}
  {021026} (\bibinfo {year} {2017})}\BibitemShut {NoStop}%
\bibitem [{\citenamefont {Xiao}\ \emph {et~al.}(2012)\citenamefont {Xiao},
  \citenamefont {Liu}, \citenamefont {Feng}, \citenamefont {Xu},\ and\
  \citenamefont {Yao}}]{xiao2012coupled}%
  \BibitemOpen
  \bibfield  {author} {\bibinfo {author} {\bibfnamefont {D.}~\bibnamefont
  {Xiao}}, \bibinfo {author} {\bibfnamefont {G.-B.}\ \bibnamefont {Liu}},
  \bibinfo {author} {\bibfnamefont {W.}~\bibnamefont {Feng}}, \bibinfo {author}
  {\bibfnamefont {X.}~\bibnamefont {Xu}},\ and\ \bibinfo {author}
  {\bibfnamefont {W.}~\bibnamefont {Yao}},\ }\bibfield  {title} {\bibinfo
  {title} {{Coupled Spin and Valley Physics in Monolayers of
  ${\mathrm{MoS}}_{2}$ and Other Group-VI Dichalcogenides}},\ }\href
  {https://doi.org/10.1103/PhysRevLett.108.196802} {\bibfield  {journal}
  {\bibinfo  {journal} {Phys. Rev. Lett.}\ }\textbf {\bibinfo {volume} {108}},\
  \bibinfo {pages} {196802} (\bibinfo {year} {2012})}\BibitemShut {NoStop}%
\bibitem [{\citenamefont {Wilson}\ and\ \citenamefont
  {Yoffe}(1969)}]{wilson1969}%
  \BibitemOpen
  \bibfield  {author} {\bibinfo {author} {\bibfnamefont {J.}~\bibnamefont
  {Wilson}}\ and\ \bibinfo {author} {\bibfnamefont {A.}~\bibnamefont {Yoffe}},\
  }\bibfield  {title} {\bibinfo {title} {{The transition metal dichalcogenides
  discussion and interpretation of the observed optical, electrical and
  structural properties}},\ }\href {https://doi.org/10.1080/00018736900101307}
  {\bibfield  {journal} {\bibinfo  {journal} {Advances in Physics}\ }\textbf
  {\bibinfo {volume} {18}},\ \bibinfo {pages} {193} (\bibinfo {year} {1969})},\
  \Eprint {https://arxiv.org/abs/https://doi.org/10.1080/00018736900101307}
  {https://doi.org/10.1080/00018736900101307} \BibitemShut {NoStop}%
\bibitem [{\citenamefont {Beal}\ \emph {et~al.}(1972)\citenamefont {Beal},
  \citenamefont {Knights},\ and\ \citenamefont {Liang}}]{beal1972transmission}%
  \BibitemOpen
  \bibfield  {author} {\bibinfo {author} {\bibfnamefont {A.~R.}\ \bibnamefont
  {Beal}}, \bibinfo {author} {\bibfnamefont {J.~C.}\ \bibnamefont {Knights}},\
  and\ \bibinfo {author} {\bibfnamefont {W.~Y.}\ \bibnamefont {Liang}},\
  }\bibfield  {title} {\bibinfo {title} {{Transmission spectra of some
  transition metal dichalcogenides. II. Group VIA: trigonal prismatic
  coordination}},\ }\href {https://doi.org/10.1088/0022-3719/5/24/016}
  {\bibfield  {journal} {\bibinfo  {journal} {Journal of Physics C: Solid State
  Physics}\ }\textbf {\bibinfo {volume} {5}},\ \bibinfo {pages} {3540}
  (\bibinfo {year} {1972})}\BibitemShut {NoStop}%
\bibitem [{\citenamefont {Zhao}\ \emph {et~al.}(2013)\citenamefont {Zhao},
  \citenamefont {Ghorannevis}, \citenamefont {Chu}, \citenamefont {Toh},
  \citenamefont {Kloc}, \citenamefont {Tan},\ and\ \citenamefont
  {Eda}}]{zhao2013}%
  \BibitemOpen
  \bibfield  {author} {\bibinfo {author} {\bibfnamefont {W.}~\bibnamefont
  {Zhao}}, \bibinfo {author} {\bibfnamefont {Z.}~\bibnamefont {Ghorannevis}},
  \bibinfo {author} {\bibfnamefont {L.}~\bibnamefont {Chu}}, \bibinfo {author}
  {\bibfnamefont {M.}~\bibnamefont {Toh}}, \bibinfo {author} {\bibfnamefont
  {C.}~\bibnamefont {Kloc}}, \bibinfo {author} {\bibfnamefont {P.-H.}\
  \bibnamefont {Tan}},\ and\ \bibinfo {author} {\bibfnamefont {G.}~\bibnamefont
  {Eda}},\ }\bibfield  {title} {\bibinfo {title} {{Evolution of Electronic
  Structure in Atomically Thin Sheets of WS${}_2$ and WSe${}_2$}},\ }\href
  {https://doi.org/10.1021/nn305275h} {\bibfield  {journal} {\bibinfo
  {journal} {ACS Nano}\ }\textbf {\bibinfo {volume} {7}},\ \bibinfo {pages}
  {791} (\bibinfo {year} {2013})},\ \bibinfo {note} {pMID: 23256505},\ \Eprint
  {https://arxiv.org/abs/https://doi.org/10.1021/nn305275h}
  {https://doi.org/10.1021/nn305275h} \BibitemShut {NoStop}%
\bibitem [{\citenamefont {Li}\ \emph {et~al.}(2014{\natexlab{a}})\citenamefont
  {Li}, \citenamefont {Chernikov}, \citenamefont {Zhang}, \citenamefont
  {Rigosi}, \citenamefont {Hill}, \citenamefont {van~der Zande}, \citenamefont
  {Chenet}, \citenamefont {Shih}, \citenamefont {Hone},\ and\ \citenamefont
  {Heinz}}]{li2014measurement}%
  \BibitemOpen
  \bibfield  {author} {\bibinfo {author} {\bibfnamefont {Y.}~\bibnamefont
  {Li}}, \bibinfo {author} {\bibfnamefont {A.}~\bibnamefont {Chernikov}},
  \bibinfo {author} {\bibfnamefont {X.}~\bibnamefont {Zhang}}, \bibinfo
  {author} {\bibfnamefont {A.}~\bibnamefont {Rigosi}}, \bibinfo {author}
  {\bibfnamefont {H.~M.}\ \bibnamefont {Hill}}, \bibinfo {author}
  {\bibfnamefont {A.~M.}\ \bibnamefont {van~der Zande}}, \bibinfo {author}
  {\bibfnamefont {D.~A.}\ \bibnamefont {Chenet}}, \bibinfo {author}
  {\bibfnamefont {E.-M.}\ \bibnamefont {Shih}}, \bibinfo {author}
  {\bibfnamefont {J.}~\bibnamefont {Hone}},\ and\ \bibinfo {author}
  {\bibfnamefont {T.~F.}\ \bibnamefont {Heinz}},\ }\bibfield  {title} {\bibinfo
  {title} {{Measurement of the optical dielectric function of monolayer
  transition-metal dichalcogenides: ${\mathrm{MoS}}_{2}$,
  $\mathrm{Mo}\mathrm{S}{\mathrm{e}}_{2}$, ${\mathrm{WS}}_{2}$, and
  $\mathrm{WS}{\mathrm{e}}_{2}$}},\ }\href
  {https://doi.org/10.1103/PhysRevB.90.205422} {\bibfield  {journal} {\bibinfo
  {journal} {Phys. Rev. B}\ }\textbf {\bibinfo {volume} {90}},\ \bibinfo
  {pages} {205422} (\bibinfo {year} {2014}{\natexlab{a}})}\BibitemShut
  {NoStop}%
\bibitem [{\citenamefont {Li}\ \emph {et~al.}(2014{\natexlab{b}})\citenamefont
  {Li}, \citenamefont {Birdwell}, \citenamefont {Amani}, \citenamefont {Burke},
  \citenamefont {Ling}, \citenamefont {Lee}, \citenamefont {Liang},
  \citenamefont {Peng}, \citenamefont {Richter}, \citenamefont {Kong},
  \citenamefont {Gundlach},\ and\ \citenamefont {Nguyen}}]{li2014broadband}%
  \BibitemOpen
  \bibfield  {author} {\bibinfo {author} {\bibfnamefont {W.}~\bibnamefont
  {Li}}, \bibinfo {author} {\bibfnamefont {A.~G.}\ \bibnamefont {Birdwell}},
  \bibinfo {author} {\bibfnamefont {M.}~\bibnamefont {Amani}}, \bibinfo
  {author} {\bibfnamefont {R.~A.}\ \bibnamefont {Burke}}, \bibinfo {author}
  {\bibfnamefont {X.}~\bibnamefont {Ling}}, \bibinfo {author} {\bibfnamefont
  {Y.-H.}\ \bibnamefont {Lee}}, \bibinfo {author} {\bibfnamefont
  {X.}~\bibnamefont {Liang}}, \bibinfo {author} {\bibfnamefont
  {L.}~\bibnamefont {Peng}}, \bibinfo {author} {\bibfnamefont {C.~A.}\
  \bibnamefont {Richter}}, \bibinfo {author} {\bibfnamefont {J.}~\bibnamefont
  {Kong}}, \bibinfo {author} {\bibfnamefont {D.~J.}\ \bibnamefont {Gundlach}},\
  and\ \bibinfo {author} {\bibfnamefont {N.~V.}\ \bibnamefont {Nguyen}},\
  }\bibfield  {title} {\bibinfo {title} {{Broadband optical properties of
  large-area monolayer CVD molybdenum disulfide}},\ }\href
  {https://doi.org/10.1103/PhysRevB.90.195434} {\bibfield  {journal} {\bibinfo
  {journal} {Phys. Rev. B}\ }\textbf {\bibinfo {volume} {90}},\ \bibinfo
  {pages} {195434} (\bibinfo {year} {2014}{\natexlab{b}})}\BibitemShut
  {NoStop}%
\bibitem [{\citenamefont {Song}\ \emph {et~al.}(2019)\citenamefont {Song},
  \citenamefont {Gu}, \citenamefont {Fang}, \citenamefont {Ho}, \citenamefont
  {Chen}, \citenamefont {Jiang},\ and\ \citenamefont {Liu}}]{song2019complex}%
  \BibitemOpen
  \bibfield  {author} {\bibinfo {author} {\bibfnamefont {B.}~\bibnamefont
  {Song}}, \bibinfo {author} {\bibfnamefont {H.}~\bibnamefont {Gu}}, \bibinfo
  {author} {\bibfnamefont {M.}~\bibnamefont {Fang}}, \bibinfo {author}
  {\bibfnamefont {Y.-T.}\ \bibnamefont {Ho}}, \bibinfo {author} {\bibfnamefont
  {X.}~\bibnamefont {Chen}}, \bibinfo {author} {\bibfnamefont {H.}~\bibnamefont
  {Jiang}},\ and\ \bibinfo {author} {\bibfnamefont {S.}~\bibnamefont {Liu}},\
  }\bibfield  {title} {\bibinfo {title} {{Complex Optical Conductivity of
  Two-Dimensional MoS${}_2$: A Striking Layer Dependency}},\ }\href
  {https://doi.org/10.1021/acs.jpclett.9b02111} {\bibfield  {journal} {\bibinfo
   {journal} {The Journal of Physical Chemistry Letters}\ }\textbf {\bibinfo
  {volume} {10}},\ \bibinfo {pages} {6246} (\bibinfo {year} {2019})},\ \bibinfo
  {note} {pMID: 31490075},\ \Eprint
  {https://arxiv.org/abs/https://doi.org/10.1021/acs.jpclett.9b02111}
  {https://doi.org/10.1021/acs.jpclett.9b02111} \BibitemShut {NoStop}%
\bibitem [{\citenamefont {Ermolaev}\ \emph {et~al.}(2020)\citenamefont
  {Ermolaev}, \citenamefont {Stebunov}, \citenamefont {Vyshnevyy},
  \citenamefont {Tatarkin}, \citenamefont {Yakubovsky}, \citenamefont
  {Novikov}, \citenamefont {Baranov}, \citenamefont {Shegai}, \citenamefont
  {Nikitin}, \citenamefont {Arsenin},\ and\ \citenamefont
  {Volkov}}]{ermolaev2020broadband}%
  \BibitemOpen
  \bibfield  {author} {\bibinfo {author} {\bibfnamefont {G.~A.}\ \bibnamefont
  {Ermolaev}}, \bibinfo {author} {\bibfnamefont {Y.~V.}\ \bibnamefont
  {Stebunov}}, \bibinfo {author} {\bibfnamefont {A.~A.}\ \bibnamefont
  {Vyshnevyy}}, \bibinfo {author} {\bibfnamefont {D.~E.}\ \bibnamefont
  {Tatarkin}}, \bibinfo {author} {\bibfnamefont {D.~I.}\ \bibnamefont
  {Yakubovsky}}, \bibinfo {author} {\bibfnamefont {S.~M.}\ \bibnamefont
  {Novikov}}, \bibinfo {author} {\bibfnamefont {D.~G.}\ \bibnamefont
  {Baranov}}, \bibinfo {author} {\bibfnamefont {T.}~\bibnamefont {Shegai}},
  \bibinfo {author} {\bibfnamefont {A.~Y.}\ \bibnamefont {Nikitin}}, \bibinfo
  {author} {\bibfnamefont {A.~V.}\ \bibnamefont {Arsenin}},\ and\ \bibinfo
  {author} {\bibfnamefont {V.~S.}\ \bibnamefont {Volkov}},\ }\bibfield  {title}
  {\bibinfo {title} {{Broadband optical properties of monolayer and bulk
  MoS${}_2$}},\ }\href {https://doi.org/10.1038/s41699-020-0155-x} {\bibfield
  {journal} {\bibinfo  {journal} {npj 2D Materials and Applications}\ }\textbf
  {\bibinfo {volume} {4}},\ \bibinfo {pages} {21} (\bibinfo {year}
  {2020})}\BibitemShut {NoStop}%
\bibitem [{\citenamefont {Ermolaev}\ \emph {et~al.}(2021)\citenamefont
  {Ermolaev}, \citenamefont {Yakubovsky}, \citenamefont {Stebunov},
  \citenamefont {Voronov}, \citenamefont {Arsenin},\ and\ \citenamefont
  {Volkov}}]{ermolaev2021spectroscopic}%
  \BibitemOpen
  \bibfield  {author} {\bibinfo {author} {\bibfnamefont {G.~A.}\ \bibnamefont
  {Ermolaev}}, \bibinfo {author} {\bibfnamefont {D.~I.}\ \bibnamefont
  {Yakubovsky}}, \bibinfo {author} {\bibfnamefont {Y.~V.}\ \bibnamefont
  {Stebunov}}, \bibinfo {author} {\bibfnamefont {A.~A.}\ \bibnamefont
  {Voronov}}, \bibinfo {author} {\bibfnamefont {A.~V.}\ \bibnamefont
  {Arsenin}},\ and\ \bibinfo {author} {\bibfnamefont {V.~S.}\ \bibnamefont
  {Volkov}},\ }\bibfield  {title} {\bibinfo {title} {{Spectroscopic
  ellipsometry of large area monolayer WS${}_2$ and WSe${}_2$ films}},\ }\href
  {https://doi.org/10.1063/5.0054947} {\bibfield  {journal} {\bibinfo
  {journal} {AIP Conference Proceedings}\ }\textbf {\bibinfo {volume} {2359}},\
  \bibinfo {pages} {020005} (\bibinfo {year} {2021})}\BibitemShut {NoStop}%
\bibitem [{\citenamefont {Ramasubramaniam}(2012)}]{ramasubramaniam2012large}%
  \BibitemOpen
  \bibfield  {author} {\bibinfo {author} {\bibfnamefont {A.}~\bibnamefont
  {Ramasubramaniam}},\ }\bibfield  {title} {\bibinfo {title} {Large excitonic
  effects in monolayers of molybdenum and tungsten dichalcogenides},\ }\href
  {https://doi.org/10.1103/PhysRevB.86.115409} {\bibfield  {journal} {\bibinfo
  {journal} {Phys. Rev. B}\ }\textbf {\bibinfo {volume} {86}},\ \bibinfo
  {pages} {115409} (\bibinfo {year} {2012})}\BibitemShut {NoStop}%
\bibitem [{\citenamefont {Qiu}\ \emph {et~al.}(2013)\citenamefont {Qiu},
  \citenamefont {da~Jornada},\ and\ \citenamefont {Louie}}]{qiu2013optical}%
  \BibitemOpen
  \bibfield  {author} {\bibinfo {author} {\bibfnamefont {D.~Y.}\ \bibnamefont
  {Qiu}}, \bibinfo {author} {\bibfnamefont {F.~H.}\ \bibnamefont
  {da~Jornada}},\ and\ \bibinfo {author} {\bibfnamefont {S.~G.}\ \bibnamefont
  {Louie}},\ }\bibfield  {title} {\bibinfo {title} {{Optical Spectrum of
  ${\mathrm{MoS}}_{2}$: Many-Body Effects and Diversity of Exciton States}},\
  }\href {https://doi.org/10.1103/PhysRevLett.111.216805} {\bibfield  {journal}
  {\bibinfo  {journal} {Phys. Rev. Lett.}\ }\textbf {\bibinfo {volume} {111}},\
  \bibinfo {pages} {216805} (\bibinfo {year} {2013})}\BibitemShut {NoStop}%
\bibitem [{\citenamefont {Schmidt}\ \emph {et~al.}(2016)\citenamefont
  {Schmidt}, \citenamefont {Niehues}, \citenamefont {Schneider}, \citenamefont
  {Drüppel}, \citenamefont {Deilmann}, \citenamefont {Rohlfing}, \citenamefont
  {de~Vasconcellos}, \citenamefont {Castellanos-Gomez},\ and\ \citenamefont
  {Bratschitsch}}]{schmidt2016reversible}%
  \BibitemOpen
  \bibfield  {author} {\bibinfo {author} {\bibfnamefont {R.}~\bibnamefont
  {Schmidt}}, \bibinfo {author} {\bibfnamefont {I.}~\bibnamefont {Niehues}},
  \bibinfo {author} {\bibfnamefont {R.}~\bibnamefont {Schneider}}, \bibinfo
  {author} {\bibfnamefont {M.}~\bibnamefont {Drüppel}}, \bibinfo {author}
  {\bibfnamefont {T.}~\bibnamefont {Deilmann}}, \bibinfo {author}
  {\bibfnamefont {M.}~\bibnamefont {Rohlfing}}, \bibinfo {author}
  {\bibfnamefont {S.~M.}\ \bibnamefont {de~Vasconcellos}}, \bibinfo {author}
  {\bibfnamefont {A.}~\bibnamefont {Castellanos-Gomez}},\ and\ \bibinfo
  {author} {\bibfnamefont {R.}~\bibnamefont {Bratschitsch}},\ }\bibfield
  {title} {\bibinfo {title} {{Reversible uniaxial strain tuning in atomically
  thin WSe${}_2$}},\ }\href {https://doi.org/10.1088/2053-1583/3/2/021011}
  {\bibfield  {journal} {\bibinfo  {journal} {2D Materials}\ }\textbf {\bibinfo
  {volume} {3}},\ \bibinfo {pages} {021011} (\bibinfo {year}
  {2016})}\BibitemShut {NoStop}%
\bibitem [{\citenamefont {Gu}\ \emph {et~al.}(2019)\citenamefont {Gu},
  \citenamefont {Song}, \citenamefont {Fang}, \citenamefont {Hong},
  \citenamefont {Chen}, \citenamefont {Jiang}, \citenamefont {Ren},\ and\
  \citenamefont {Liu}}]{gu2019layer}%
  \BibitemOpen
  \bibfield  {author} {\bibinfo {author} {\bibfnamefont {H.}~\bibnamefont
  {Gu}}, \bibinfo {author} {\bibfnamefont {B.}~\bibnamefont {Song}}, \bibinfo
  {author} {\bibfnamefont {M.}~\bibnamefont {Fang}}, \bibinfo {author}
  {\bibfnamefont {Y.}~\bibnamefont {Hong}}, \bibinfo {author} {\bibfnamefont
  {X.}~\bibnamefont {Chen}}, \bibinfo {author} {\bibfnamefont {H.}~\bibnamefont
  {Jiang}}, \bibinfo {author} {\bibfnamefont {W.}~\bibnamefont {Ren}},\ and\
  \bibinfo {author} {\bibfnamefont {S.}~\bibnamefont {Liu}},\ }\bibfield
  {title} {\bibinfo {title} {{Layer-dependent dielectric and optical properties
  of centimeter-scale 2D WSe${}_2$: evolution from a single layer to few
  layers}},\ }\href {https://doi.org/10.1039/C9NR04270A} {\bibfield  {journal}
  {\bibinfo  {journal} {Nanoscale}\ }\textbf {\bibinfo {volume} {11}},\
  \bibinfo {pages} {22762} (\bibinfo {year} {2019})}\BibitemShut {NoStop}%
\bibitem [{\citenamefont {Woo}\ \emph {et~al.}(2023)\citenamefont {Woo},
  \citenamefont {Zobelli}, \citenamefont {Schneider}, \citenamefont {Arora},
  \citenamefont {Preu\ss{}}, \citenamefont {Carey}, \citenamefont {Michaelis~de
  Vasconcellos}, \citenamefont {Palummo}, \citenamefont {Bratschitsch},\ and\
  \citenamefont {Tizei}}]{woo2023excitonic}%
  \BibitemOpen
  \bibfield  {author} {\bibinfo {author} {\bibfnamefont {S.~Y.}\ \bibnamefont
  {Woo}}, \bibinfo {author} {\bibfnamefont {A.}~\bibnamefont {Zobelli}},
  \bibinfo {author} {\bibfnamefont {R.}~\bibnamefont {Schneider}}, \bibinfo
  {author} {\bibfnamefont {A.}~\bibnamefont {Arora}}, \bibinfo {author}
  {\bibfnamefont {J.~A.}\ \bibnamefont {Preu\ss{}}}, \bibinfo {author}
  {\bibfnamefont {B.~J.}\ \bibnamefont {Carey}}, \bibinfo {author}
  {\bibfnamefont {S.}~\bibnamefont {Michaelis~de Vasconcellos}}, \bibinfo
  {author} {\bibfnamefont {M.}~\bibnamefont {Palummo}}, \bibinfo {author}
  {\bibfnamefont {R.}~\bibnamefont {Bratschitsch}},\ and\ \bibinfo {author}
  {\bibfnamefont {L.~H.~G.}\ \bibnamefont {Tizei}},\ }\bibfield  {title}
  {\bibinfo {title} {{Excitonic absorption signatures of twisted bilayer
  ${\mathrm{WSe}}_{2}$ by electron energy-loss spectroscopy}},\ }\href
  {https://doi.org/10.1103/PhysRevB.107.155429} {\bibfield  {journal} {\bibinfo
   {journal} {Phys. Rev. B}\ }\textbf {\bibinfo {volume} {107}},\ \bibinfo
  {pages} {155429} (\bibinfo {year} {2023})}\BibitemShut {NoStop}%
\bibitem [{\citenamefont {Aivazian}\ \emph {et~al.}(2017)\citenamefont
  {Aivazian}, \citenamefont {Yu}, \citenamefont {Wu}, \citenamefont {Yan},
  \citenamefont {Mandrus}, \citenamefont {Cobden}, \citenamefont {Yao},\ and\
  \citenamefont {Xu}}]{aivazian2017manybody}%
  \BibitemOpen
  \bibfield  {author} {\bibinfo {author} {\bibfnamefont {G.}~\bibnamefont
  {Aivazian}}, \bibinfo {author} {\bibfnamefont {H.}~\bibnamefont {Yu}},
  \bibinfo {author} {\bibfnamefont {S.}~\bibnamefont {Wu}}, \bibinfo {author}
  {\bibfnamefont {J.}~\bibnamefont {Yan}}, \bibinfo {author} {\bibfnamefont
  {D.~G.}\ \bibnamefont {Mandrus}}, \bibinfo {author} {\bibfnamefont
  {D.}~\bibnamefont {Cobden}}, \bibinfo {author} {\bibfnamefont
  {W.}~\bibnamefont {Yao}},\ and\ \bibinfo {author} {\bibfnamefont
  {X.}~\bibnamefont {Xu}},\ }\bibfield  {title} {\bibinfo {title} {{Many-body
  effects in nonlinear optical responses of 2D layered semiconductors}},\
  }\href {https://doi.org/10.1088/2053-1583/aa56f1} {\bibfield  {journal}
  {\bibinfo  {journal} {2D Materials}\ }\textbf {\bibinfo {volume} {4}},\
  \bibinfo {pages} {025024} (\bibinfo {year} {2017})}\BibitemShut {NoStop}%
\bibitem [{\citenamefont {Trovatello}\ \emph {et~al.}(2020)\citenamefont
  {Trovatello}, \citenamefont {Katsch}, \citenamefont {Borys}, \citenamefont
  {Selig}, \citenamefont {Yao}, \citenamefont {Borrego-Varillas}, \citenamefont
  {Scotognella}, \citenamefont {Kriegel}, \citenamefont {Yan}, \citenamefont
  {Zettl}, \citenamefont {Schuck}, \citenamefont {Knorr}, \citenamefont
  {Cerullo},\ and\ \citenamefont {Conte}}]{trovatello2020ultrafast}%
  \BibitemOpen
  \bibfield  {author} {\bibinfo {author} {\bibfnamefont {C.}~\bibnamefont
  {Trovatello}}, \bibinfo {author} {\bibfnamefont {F.}~\bibnamefont {Katsch}},
  \bibinfo {author} {\bibfnamefont {N.~J.}\ \bibnamefont {Borys}}, \bibinfo
  {author} {\bibfnamefont {M.}~\bibnamefont {Selig}}, \bibinfo {author}
  {\bibfnamefont {K.}~\bibnamefont {Yao}}, \bibinfo {author} {\bibfnamefont
  {R.}~\bibnamefont {Borrego-Varillas}}, \bibinfo {author} {\bibfnamefont
  {F.}~\bibnamefont {Scotognella}}, \bibinfo {author} {\bibfnamefont
  {I.}~\bibnamefont {Kriegel}}, \bibinfo {author} {\bibfnamefont
  {A.}~\bibnamefont {Yan}}, \bibinfo {author} {\bibfnamefont {A.}~\bibnamefont
  {Zettl}}, \bibinfo {author} {\bibfnamefont {P.~J.}\ \bibnamefont {Schuck}},
  \bibinfo {author} {\bibfnamefont {A.}~\bibnamefont {Knorr}}, \bibinfo
  {author} {\bibfnamefont {G.}~\bibnamefont {Cerullo}},\ and\ \bibinfo {author}
  {\bibfnamefont {S.~D.}\ \bibnamefont {Conte}},\ }\bibfield  {title} {\bibinfo
  {title} {{The ultrafast onset of exciton formation in 2D semiconductors}},\
  }\href {https://doi.org/10.1038/s41467-020-18835-5} {\bibfield  {journal}
  {\bibinfo  {journal} {Nature Communications}\ }\textbf {\bibinfo {volume}
  {11}},\ \bibinfo {pages} {5277} (\bibinfo {year} {2020})}\BibitemShut
  {NoStop}%
\bibitem [{\citenamefont {Genco}\ \emph {et~al.}(2023)\citenamefont {Genco},
  \citenamefont {Trovatello}, \citenamefont {Louca}, \citenamefont {Watanabe},
  \citenamefont {Taniguchi}, \citenamefont {Tartakovskii}, \citenamefont
  {Cerullo},\ and\ \citenamefont {Dal~Conte}}]{genco2023ultrafast}%
  \BibitemOpen
  \bibfield  {author} {\bibinfo {author} {\bibfnamefont {A.}~\bibnamefont
  {Genco}}, \bibinfo {author} {\bibfnamefont {C.}~\bibnamefont {Trovatello}},
  \bibinfo {author} {\bibfnamefont {C.}~\bibnamefont {Louca}}, \bibinfo
  {author} {\bibfnamefont {K.}~\bibnamefont {Watanabe}}, \bibinfo {author}
  {\bibfnamefont {T.}~\bibnamefont {Taniguchi}}, \bibinfo {author}
  {\bibfnamefont {A.~I.}\ \bibnamefont {Tartakovskii}}, \bibinfo {author}
  {\bibfnamefont {G.}~\bibnamefont {Cerullo}},\ and\ \bibinfo {author}
  {\bibfnamefont {S.}~\bibnamefont {Dal~Conte}},\ }\bibfield  {title} {\bibinfo
  {title} {{Ultrafast Exciton and Trion Dynamics in High-Quality Encapsulated
  MoS${}_2$ Monolayers}},\ }\href
  {https://doi.org/https://doi.org/10.1002/pssb.202200376} {\bibfield
  {journal} {\bibinfo  {journal} {physica status solidi (b)}\ }\textbf
  {\bibinfo {volume} {260}},\ \bibinfo {pages} {2200376} (\bibinfo {year}
  {2023})}\BibitemShut {NoStop}%
\bibitem [{\citenamefont {Genco}\ \emph {et~al.}(2025)\citenamefont {Genco},
  \citenamefont {Trovatello}, \citenamefont {Shahnazaryan}, \citenamefont
  {Dogadov}, \citenamefont {Cadore}, \citenamefont {Rosa}, \citenamefont
  {Kerfoot}, \citenamefont {Ahmed}, \citenamefont {Balci}, \citenamefont
  {Alexeev}, \citenamefont {Rostami}, \citenamefont {Watanabe}, \citenamefont
  {Taniguchi}, \citenamefont {Tongay}, \citenamefont {Ferrari}, \citenamefont
  {Cerullo},\ and\ \citenamefont {Dal~Conte}}]{genco2025ultrafast}%
  \BibitemOpen
  \bibfield  {author} {\bibinfo {author} {\bibfnamefont {A.}~\bibnamefont
  {Genco}}, \bibinfo {author} {\bibfnamefont {C.}~\bibnamefont {Trovatello}},
  \bibinfo {author} {\bibfnamefont {V.~A.}\ \bibnamefont {Shahnazaryan}},
  \bibinfo {author} {\bibfnamefont {O.}~\bibnamefont {Dogadov}}, \bibinfo
  {author} {\bibfnamefont {A.~R.}\ \bibnamefont {Cadore}}, \bibinfo {author}
  {\bibfnamefont {B.~L.~T.}\ \bibnamefont {Rosa}}, \bibinfo {author}
  {\bibfnamefont {J.~A.}\ \bibnamefont {Kerfoot}}, \bibinfo {author}
  {\bibfnamefont {T.}~\bibnamefont {Ahmed}}, \bibinfo {author} {\bibfnamefont
  {O.}~\bibnamefont {Balci}}, \bibinfo {author} {\bibfnamefont {E.~M.}\
  \bibnamefont {Alexeev}}, \bibinfo {author} {\bibfnamefont {H.}~\bibnamefont
  {Rostami}}, \bibinfo {author} {\bibfnamefont {K.}~\bibnamefont {Watanabe}},
  \bibinfo {author} {\bibfnamefont {T.}~\bibnamefont {Taniguchi}}, \bibinfo
  {author} {\bibfnamefont {S.~A.}\ \bibnamefont {Tongay}}, \bibinfo {author}
  {\bibfnamefont {A.~C.}\ \bibnamefont {Ferrari}}, \bibinfo {author}
  {\bibfnamefont {G.}~\bibnamefont {Cerullo}},\ and\ \bibinfo {author}
  {\bibfnamefont {S.}~\bibnamefont {Dal~Conte}},\ }\bibfield  {title} {\bibinfo
  {title} {{Ultrafast Dynamics of Rydberg Excitons and Their Optically Induced
  Charged Complexes in Encapsulated WSe${}_2$ Monolayers}},\ }\href
  {https://doi.org/10.1021/acs.nanolett.4c06428} {\bibfield  {journal}
  {\bibinfo  {journal} {Nano Letters}\ }\textbf {\bibinfo {volume} {25}},\
  \bibinfo {pages} {7673} (\bibinfo {year} {2025})},\ \bibinfo {note} {pMID:
  40305446},\ \Eprint
  {https://arxiv.org/abs/https://doi.org/10.1021/acs.nanolett.4c06428}
  {https://doi.org/10.1021/acs.nanolett.4c06428} \BibitemShut {NoStop}%
\bibitem [{\citenamefont {Nie}\ \emph {et~al.}(2019)\citenamefont {Nie},
  \citenamefont {Shi}, \citenamefont {Qin}, \citenamefont {Wang}, \citenamefont
  {Jiang}, \citenamefont {Zheng}, \citenamefont {Cui}, \citenamefont {Meng},
  \citenamefont {Song}, \citenamefont {Wang}, \citenamefont {Turcu},
  \citenamefont {Wang}, \citenamefont {Xu}, \citenamefont {Shi}, \citenamefont
  {Zhao}, \citenamefont {Zhang},\ and\ \citenamefont
  {Wang}}]{nie2019tailoring}%
  \BibitemOpen
  \bibfield  {author} {\bibinfo {author} {\bibfnamefont {Z.}~\bibnamefont
  {Nie}}, \bibinfo {author} {\bibfnamefont {Y.}~\bibnamefont {Shi}}, \bibinfo
  {author} {\bibfnamefont {S.}~\bibnamefont {Qin}}, \bibinfo {author}
  {\bibfnamefont {Y.}~\bibnamefont {Wang}}, \bibinfo {author} {\bibfnamefont
  {H.}~\bibnamefont {Jiang}}, \bibinfo {author} {\bibfnamefont
  {Q.}~\bibnamefont {Zheng}}, \bibinfo {author} {\bibfnamefont
  {Y.}~\bibnamefont {Cui}}, \bibinfo {author} {\bibfnamefont {Y.}~\bibnamefont
  {Meng}}, \bibinfo {author} {\bibfnamefont {F.}~\bibnamefont {Song}}, \bibinfo
  {author} {\bibfnamefont {X.}~\bibnamefont {Wang}}, \bibinfo {author}
  {\bibfnamefont {I.~C.~E.}\ \bibnamefont {Turcu}}, \bibinfo {author}
  {\bibfnamefont {X.}~\bibnamefont {Wang}}, \bibinfo {author} {\bibfnamefont
  {Y.}~\bibnamefont {Xu}}, \bibinfo {author} {\bibfnamefont {Y.}~\bibnamefont
  {Shi}}, \bibinfo {author} {\bibfnamefont {J.}~\bibnamefont {Zhao}}, \bibinfo
  {author} {\bibfnamefont {R.}~\bibnamefont {Zhang}},\ and\ \bibinfo {author}
  {\bibfnamefont {F.}~\bibnamefont {Wang}},\ }\bibfield  {title} {\bibinfo
  {title} {Tailoring exciton dynamics of monolayer transition metal
  dichalcogenides by interfacial electron-phonon coupling},\ }\href
  {https://doi.org/10.1038/s42005-019-0202-0} {\bibfield  {journal} {\bibinfo
  {journal} {Communications Physics}\ }\textbf {\bibinfo {volume} {2}},\
  \bibinfo {pages} {103} (\bibinfo {year} {2019})}\BibitemShut {NoStop}%
\bibitem [{\citenamefont {Morabito}\ \emph {et~al.}(2024)\citenamefont
  {Morabito}, \citenamefont {Synnatschke}, \citenamefont {Mehew}, \citenamefont
  {Varghese}, \citenamefont {Sayers}, \citenamefont {Folpini}, \citenamefont
  {Petrozza}, \citenamefont {Cerullo}, \citenamefont {Tielrooij}, \citenamefont
  {Coleman}, \citenamefont {Nicolosi},\ and\ \citenamefont
  {Gadermaier}}]{morabito2024long}%
  \BibitemOpen
  \bibfield  {author} {\bibinfo {author} {\bibfnamefont {F.}~\bibnamefont
  {Morabito}}, \bibinfo {author} {\bibfnamefont {K.}~\bibnamefont
  {Synnatschke}}, \bibinfo {author} {\bibfnamefont {J.~D.}\ \bibnamefont
  {Mehew}}, \bibinfo {author} {\bibfnamefont {S.}~\bibnamefont {Varghese}},
  \bibinfo {author} {\bibfnamefont {C.~J.}\ \bibnamefont {Sayers}}, \bibinfo
  {author} {\bibfnamefont {G.}~\bibnamefont {Folpini}}, \bibinfo {author}
  {\bibfnamefont {A.}~\bibnamefont {Petrozza}}, \bibinfo {author}
  {\bibfnamefont {G.}~\bibnamefont {Cerullo}}, \bibinfo {author} {\bibfnamefont
  {K.-J.}\ \bibnamefont {Tielrooij}}, \bibinfo {author} {\bibfnamefont
  {J.}~\bibnamefont {Coleman}}, \bibinfo {author} {\bibfnamefont
  {V.}~\bibnamefont {Nicolosi}},\ and\ \bibinfo {author} {\bibfnamefont
  {C.}~\bibnamefont {Gadermaier}},\ }\bibfield  {title} {\bibinfo {title}
  {{Long lived photogenerated charge carriers in few-layer transition metal
  dichalcogenides obtained from liquid phase exfoliation}},\ }\href
  {https://doi.org/10.1039/D3NA00862B} {\bibfield  {journal} {\bibinfo
  {journal} {Nanoscale Adv.}\ }\textbf {\bibinfo {volume} {6}},\ \bibinfo
  {pages} {1074} (\bibinfo {year} {2024})}\BibitemShut {NoStop}%
\bibitem [{\citenamefont {Steinleitner}\ \emph {et~al.}(2017)\citenamefont
  {Steinleitner}, \citenamefont {Merkl}, \citenamefont {Nagler}, \citenamefont
  {Mornhinweg}, \citenamefont {Sch{\"u}ller}, \citenamefont {Korn},
  \citenamefont {Chernikov},\ and\ \citenamefont
  {Huber}}]{steinleitner2017direct}%
  \BibitemOpen
  \bibfield  {author} {\bibinfo {author} {\bibfnamefont {P.}~\bibnamefont
  {Steinleitner}}, \bibinfo {author} {\bibfnamefont {P.}~\bibnamefont {Merkl}},
  \bibinfo {author} {\bibfnamefont {P.}~\bibnamefont {Nagler}}, \bibinfo
  {author} {\bibfnamefont {J.}~\bibnamefont {Mornhinweg}}, \bibinfo {author}
  {\bibfnamefont {C.}~\bibnamefont {Sch{\"u}ller}}, \bibinfo {author}
  {\bibfnamefont {T.}~\bibnamefont {Korn}}, \bibinfo {author} {\bibfnamefont
  {A.}~\bibnamefont {Chernikov}},\ and\ \bibinfo {author} {\bibfnamefont
  {R.}~\bibnamefont {Huber}},\ }\bibfield  {title} {\bibinfo {title} {{Direct
  Observation of Ultrafast Exciton Formation in a Monolayer of WSe${}_2$}},\
  }\href {https://doi.org/10.1021/acs.nanolett.6b04422} {\bibfield  {journal}
  {\bibinfo  {journal} {Nano Letters}\ }\textbf {\bibinfo {volume} {17}},\
  \bibinfo {pages} {1455} (\bibinfo {year} {2017})},\ \bibinfo {note} {pMID:
  28182430},\ \Eprint
  {https://arxiv.org/abs/https://doi.org/10.1021/acs.nanolett.6b04422}
  {https://doi.org/10.1021/acs.nanolett.6b04422} \BibitemShut {NoStop}%
\bibitem [{\citenamefont {Robert}\ \emph {et~al.}(2016)\citenamefont {Robert},
  \citenamefont {Lagarde}, \citenamefont {Cadiz}, \citenamefont {Wang},
  \citenamefont {Lassagne}, \citenamefont {Amand}, \citenamefont {Balocchi},
  \citenamefont {Renucci}, \citenamefont {Tongay}, \citenamefont {Urbaszek},\
  and\ \citenamefont {Marie}}]{robert2016exciton}%
  \BibitemOpen
  \bibfield  {author} {\bibinfo {author} {\bibfnamefont {C.}~\bibnamefont
  {Robert}}, \bibinfo {author} {\bibfnamefont {D.}~\bibnamefont {Lagarde}},
  \bibinfo {author} {\bibfnamefont {F.}~\bibnamefont {Cadiz}}, \bibinfo
  {author} {\bibfnamefont {G.}~\bibnamefont {Wang}}, \bibinfo {author}
  {\bibfnamefont {B.}~\bibnamefont {Lassagne}}, \bibinfo {author}
  {\bibfnamefont {T.}~\bibnamefont {Amand}}, \bibinfo {author} {\bibfnamefont
  {A.}~\bibnamefont {Balocchi}}, \bibinfo {author} {\bibfnamefont
  {P.}~\bibnamefont {Renucci}}, \bibinfo {author} {\bibfnamefont
  {S.}~\bibnamefont {Tongay}}, \bibinfo {author} {\bibfnamefont
  {B.}~\bibnamefont {Urbaszek}},\ and\ \bibinfo {author} {\bibfnamefont
  {X.}~\bibnamefont {Marie}},\ }\bibfield  {title} {\bibinfo {title} {Exciton
  radiative lifetime in transition metal dichalcogenide monolayers},\ }\href
  {https://doi.org/10.1103/PhysRevB.93.205423} {\bibfield  {journal} {\bibinfo
  {journal} {Phys. Rev. B}\ }\textbf {\bibinfo {volume} {93}},\ \bibinfo
  {pages} {205423} (\bibinfo {year} {2016})}\BibitemShut {NoStop}%
\bibitem [{\citenamefont {Dogadov}\ \emph
  {et~al.}(2026{\natexlab{a}})\citenamefont {Dogadov}, \citenamefont
  {Mittenzwey}, \citenamefont {Bertolotti}, \citenamefont {Olsen},
  \citenamefont {Deckert}, \citenamefont {Trovatello}, \citenamefont {Zhu},
  \citenamefont {Brida}, \citenamefont {Cerullo}, \citenamefont {Knorr},\ and\
  \citenamefont {Dal~Conte}}]{dogadov2025dissecting}%
  \BibitemOpen
  \bibfield  {author} {\bibinfo {author} {\bibfnamefont {O.}~\bibnamefont
  {Dogadov}}, \bibinfo {author} {\bibfnamefont {H.}~\bibnamefont {Mittenzwey}},
  \bibinfo {author} {\bibfnamefont {M.}~\bibnamefont {Bertolotti}}, \bibinfo
  {author} {\bibfnamefont {N.}~\bibnamefont {Olsen}}, \bibinfo {author}
  {\bibfnamefont {T.}~\bibnamefont {Deckert}}, \bibinfo {author} {\bibfnamefont
  {C.}~\bibnamefont {Trovatello}}, \bibinfo {author} {\bibfnamefont
  {X.}~\bibnamefont {Zhu}}, \bibinfo {author} {\bibfnamefont {D.}~\bibnamefont
  {Brida}}, \bibinfo {author} {\bibfnamefont {G.}~\bibnamefont {Cerullo}},
  \bibinfo {author} {\bibfnamefont {A.}~\bibnamefont {Knorr}},\ and\ \bibinfo
  {author} {\bibfnamefont {S.}~\bibnamefont {Dal~Conte}},\ }\bibfield  {title}
  {\bibinfo {title} {Dissecting intervalley coupling mechanisms in monolayer
  transition metal dichalcogenides},\ }\href
  {https://doi.org/10.1038/s41699-025-00653-2} {\bibfield  {journal} {\bibinfo
  {journal} {npj 2D Materials and Applications}\ }\textbf {\bibinfo {volume}
  {10}},\ \bibinfo {pages} {21} (\bibinfo {year}
  {2026}{\natexlab{a}})}\BibitemShut {NoStop}%
\bibitem [{\citenamefont {Sun}\ \emph {et~al.}(2014)\citenamefont {Sun},
  \citenamefont {Rao}, \citenamefont {Reider}, \citenamefont {Chen},
  \citenamefont {You}, \citenamefont {Br{\'e}zin}, \citenamefont
  {Harutyunyan},\ and\ \citenamefont {Heinz}}]{sun2014observation}%
  \BibitemOpen
  \bibfield  {author} {\bibinfo {author} {\bibfnamefont {D.}~\bibnamefont
  {Sun}}, \bibinfo {author} {\bibfnamefont {Y.}~\bibnamefont {Rao}}, \bibinfo
  {author} {\bibfnamefont {G.~A.}\ \bibnamefont {Reider}}, \bibinfo {author}
  {\bibfnamefont {G.}~\bibnamefont {Chen}}, \bibinfo {author} {\bibfnamefont
  {Y.}~\bibnamefont {You}}, \bibinfo {author} {\bibfnamefont {L.}~\bibnamefont
  {Br{\'e}zin}}, \bibinfo {author} {\bibfnamefont {A.~R.}\ \bibnamefont
  {Harutyunyan}},\ and\ \bibinfo {author} {\bibfnamefont {T.~F.}\ \bibnamefont
  {Heinz}},\ }\bibfield  {title} {\bibinfo {title} {{Observation of Rapid
  Exciton–Exciton Annihilation in Monolayer Molybdenum Disulfide}},\ }\href
  {https://doi.org/10.1021/nl5021975} {\bibfield  {journal} {\bibinfo
  {journal} {Nano Letters}\ }\textbf {\bibinfo {volume} {14}},\ \bibinfo
  {pages} {5625} (\bibinfo {year} {2014})},\ \bibinfo {note} {pMID: 25171389},\
  \Eprint {https://arxiv.org/abs/https://doi.org/10.1021/nl5021975}
  {https://doi.org/10.1021/nl5021975} \BibitemShut {NoStop}%
\bibitem [{\citenamefont {Cunningham}\ \emph {et~al.}(2017)\citenamefont
  {Cunningham}, \citenamefont {Hanbicki}, \citenamefont {McCreary},\ and\
  \citenamefont {Jonker}}]{cunningham2017photoinduced}%
  \BibitemOpen
  \bibfield  {author} {\bibinfo {author} {\bibfnamefont {P.~D.}\ \bibnamefont
  {Cunningham}}, \bibinfo {author} {\bibfnamefont {A.~T.}\ \bibnamefont
  {Hanbicki}}, \bibinfo {author} {\bibfnamefont {K.~M.}\ \bibnamefont
  {McCreary}},\ and\ \bibinfo {author} {\bibfnamefont {B.~T.}\ \bibnamefont
  {Jonker}},\ }\bibfield  {title} {\bibinfo {title} {{Photoinduced Bandgap
  Renormalization and Exciton Binding Energy Reduction in WS${}_2$}},\ }\href
  {https://doi.org/10.1021/acsnano.7b06885} {\bibfield  {journal} {\bibinfo
  {journal} {ACS Nano}\ }\textbf {\bibinfo {volume} {11}},\ \bibinfo {pages}
  {12601} (\bibinfo {year} {2017})},\ \bibinfo {note} {pMID: 29227085},\
  \Eprint {https://arxiv.org/abs/https://doi.org/10.1021/acsnano.7b06885}
  {https://doi.org/10.1021/acsnano.7b06885} \BibitemShut {NoStop}%
\bibitem [{\citenamefont {Pogna}\ \emph {et~al.}(2016)\citenamefont {Pogna},
  \citenamefont {Marsili}, \citenamefont {De~Fazio}, \citenamefont {Dal~Conte},
  \citenamefont {Manzoni}, \citenamefont {Sangalli}, \citenamefont {Yoon},
  \citenamefont {Lombardo}, \citenamefont {Ferrari}, \citenamefont {Marini},
  \citenamefont {Cerullo},\ and\ \citenamefont
  {Prezzi}}]{pogna2016photoinduced}%
  \BibitemOpen
  \bibfield  {author} {\bibinfo {author} {\bibfnamefont {E.~A.~A.}\
  \bibnamefont {Pogna}}, \bibinfo {author} {\bibfnamefont {M.}~\bibnamefont
  {Marsili}}, \bibinfo {author} {\bibfnamefont {D.}~\bibnamefont {De~Fazio}},
  \bibinfo {author} {\bibfnamefont {S.}~\bibnamefont {Dal~Conte}}, \bibinfo
  {author} {\bibfnamefont {C.}~\bibnamefont {Manzoni}}, \bibinfo {author}
  {\bibfnamefont {D.}~\bibnamefont {Sangalli}}, \bibinfo {author}
  {\bibfnamefont {D.}~\bibnamefont {Yoon}}, \bibinfo {author} {\bibfnamefont
  {A.}~\bibnamefont {Lombardo}}, \bibinfo {author} {\bibfnamefont {A.~C.}\
  \bibnamefont {Ferrari}}, \bibinfo {author} {\bibfnamefont {A.}~\bibnamefont
  {Marini}}, \bibinfo {author} {\bibfnamefont {G.}~\bibnamefont {Cerullo}},\
  and\ \bibinfo {author} {\bibfnamefont {D.}~\bibnamefont {Prezzi}},\
  }\bibfield  {title} {\bibinfo {title} {{Photo-Induced Bandgap Renormalization
  Governs the Ultrafast Response of Single-Layer MoS${}_2$}},\ }\href
  {https://doi.org/10.1021/acsnano.5b06488} {\bibfield  {journal} {\bibinfo
  {journal} {ACS Nano}\ }\textbf {\bibinfo {volume} {10}},\ \bibinfo {pages}
  {1182} (\bibinfo {year} {2016})},\ \bibinfo {note} {pMID: 26691058},\ \Eprint
  {https://arxiv.org/abs/https://doi.org/10.1021/acsnano.5b06488}
  {https://doi.org/10.1021/acsnano.5b06488} \BibitemShut {NoStop}%
\bibitem [{\citenamefont {Trovatello}\ \emph {et~al.}(2022)\citenamefont
  {Trovatello}, \citenamefont {Katsch}, \citenamefont {Li}, \citenamefont
  {Zhu}, \citenamefont {Knorr}, \citenamefont {Cerullo},\ and\ \citenamefont
  {Dal~Conte}}]{trovatello2022disentangling}%
  \BibitemOpen
  \bibfield  {author} {\bibinfo {author} {\bibfnamefont {C.}~\bibnamefont
  {Trovatello}}, \bibinfo {author} {\bibfnamefont {F.}~\bibnamefont {Katsch}},
  \bibinfo {author} {\bibfnamefont {Q.}~\bibnamefont {Li}}, \bibinfo {author}
  {\bibfnamefont {X.}~\bibnamefont {Zhu}}, \bibinfo {author} {\bibfnamefont
  {A.}~\bibnamefont {Knorr}}, \bibinfo {author} {\bibfnamefont
  {G.}~\bibnamefont {Cerullo}},\ and\ \bibinfo {author} {\bibfnamefont
  {S.}~\bibnamefont {Dal~Conte}},\ }\bibfield  {title} {\bibinfo {title}
  {{Disentangling Many-Body Effects in the Coherent Optical Response of 2D
  Semiconductors}},\ }\href {https://doi.org/10.1021/acs.nanolett.2c01309}
  {\bibfield  {journal} {\bibinfo  {journal} {Nano Letters}\ }\textbf {\bibinfo
  {volume} {22}},\ \bibinfo {pages} {5322} (\bibinfo {year}
  {2022})}\BibitemShut {NoStop}%
\bibitem [{\citenamefont {Deckert}\ \emph {et~al.}(2025)\citenamefont
  {Deckert}, \citenamefont {Mittenzwey}, \citenamefont {Dogadov}, \citenamefont
  {Bertolotti}, \citenamefont {Cerullo}, \citenamefont {Brida}, \citenamefont
  {Knorr},\ and\ \citenamefont {Dal~Conte}}]{deckert2025coherent}%
  \BibitemOpen
  \bibfield  {author} {\bibinfo {author} {\bibfnamefont {T.}~\bibnamefont
  {Deckert}}, \bibinfo {author} {\bibfnamefont {H.}~\bibnamefont {Mittenzwey}},
  \bibinfo {author} {\bibfnamefont {O.}~\bibnamefont {Dogadov}}, \bibinfo
  {author} {\bibfnamefont {M.}~\bibnamefont {Bertolotti}}, \bibinfo {author}
  {\bibfnamefont {G.}~\bibnamefont {Cerullo}}, \bibinfo {author} {\bibfnamefont
  {D.}~\bibnamefont {Brida}}, \bibinfo {author} {\bibfnamefont
  {A.}~\bibnamefont {Knorr}},\ and\ \bibinfo {author} {\bibfnamefont
  {S.}~\bibnamefont {Dal~Conte}},\ }\bibfield  {title} {\bibinfo {title}
  {Coherent coulomb intra- and intervalley many-body effects in single-layer
  transition metal dichalcogenides},\ }\href
  {https://doi.org/10.1103/j5cv-rffq} {\bibfield  {journal} {\bibinfo
  {journal} {Phys. Rev. Lett.}\ }\textbf {\bibinfo {volume} {135}},\ \bibinfo
  {pages} {066902} (\bibinfo {year} {2025})}\BibitemShut {NoStop}%
\bibitem [{\citenamefont {Wang}\ \emph {et~al.}(2017)\citenamefont {Wang},
  \citenamefont {Wang}, \citenamefont {Wang}, \citenamefont {Grinblat},
  \citenamefont {Huang}, \citenamefont {Wang}, \citenamefont {Ye},
  \citenamefont {Li}, \citenamefont {Bao}, \citenamefont {Wee}, \citenamefont
  {Maier}, \citenamefont {Chen}, \citenamefont {Zhong}, \citenamefont {Qiu},\
  and\ \citenamefont {Sun}}]{Wang2017}%
  \BibitemOpen
  \bibfield  {author} {\bibinfo {author} {\bibfnamefont {L.}~\bibnamefont
  {Wang}}, \bibinfo {author} {\bibfnamefont {Z.}~\bibnamefont {Wang}}, \bibinfo
  {author} {\bibfnamefont {H.-Y.}\ \bibnamefont {Wang}}, \bibinfo {author}
  {\bibfnamefont {G.}~\bibnamefont {Grinblat}}, \bibinfo {author}
  {\bibfnamefont {Y.-L.}\ \bibnamefont {Huang}}, \bibinfo {author}
  {\bibfnamefont {D.}~\bibnamefont {Wang}}, \bibinfo {author} {\bibfnamefont
  {X.-H.}\ \bibnamefont {Ye}}, \bibinfo {author} {\bibfnamefont {X.-B.}\
  \bibnamefont {Li}}, \bibinfo {author} {\bibfnamefont {Q.}~\bibnamefont
  {Bao}}, \bibinfo {author} {\bibfnamefont {A.-S.}\ \bibnamefont {Wee}},
  \bibinfo {author} {\bibfnamefont {S.~A.}\ \bibnamefont {Maier}}, \bibinfo
  {author} {\bibfnamefont {Q.-D.}\ \bibnamefont {Chen}}, \bibinfo {author}
  {\bibfnamefont {M.-L.}\ \bibnamefont {Zhong}}, \bibinfo {author}
  {\bibfnamefont {C.-W.}\ \bibnamefont {Qiu}},\ and\ \bibinfo {author}
  {\bibfnamefont {H.-B.}\ \bibnamefont {Sun}},\ }\bibfield  {title} {\bibinfo
  {title} {{Slow cooling and efficient extraction of C-exciton hot carriers in
  MoS${}_2$ monolayer}},\ }\href@noop {} {\bibfield  {journal} {\bibinfo
  {journal} {Nature Communications}\ }\textbf {\bibinfo {volume} {8}} (\bibinfo
  {year} {2017})}\BibitemShut {NoStop}%
\bibitem [{\citenamefont {Tran}\ \emph {et~al.}(2024)\citenamefont {Tran},
  \citenamefont {Jang}, \citenamefont {Vu}, \citenamefont {Jung}, \citenamefont
  {Do}, \citenamefont {Jin}, \citenamefont {Lee}, \citenamefont {Kim},\ and\
  \citenamefont {Kim}}]{tran2024augmented}%
  \BibitemOpen
  \bibfield  {author} {\bibinfo {author} {\bibfnamefont {T.-X.}\ \bibnamefont
  {Tran}}, \bibinfo {author} {\bibfnamefont {Y.~J.}\ \bibnamefont {Jang}},
  \bibinfo {author} {\bibfnamefont {V.-T.}\ \bibnamefont {Vu}}, \bibinfo
  {author} {\bibfnamefont {C.-W.}\ \bibnamefont {Jung}}, \bibinfo {author}
  {\bibfnamefont {V.~D.}\ \bibnamefont {Do}}, \bibinfo {author} {\bibfnamefont
  {Y.}~\bibnamefont {Jin}}, \bibinfo {author} {\bibfnamefont {J.}~\bibnamefont
  {Lee}}, \bibinfo {author} {\bibfnamefont {H.}~\bibnamefont {Kim}},\ and\
  \bibinfo {author} {\bibfnamefont {J.-H.}\ \bibnamefont {Kim}},\ }\bibfield
  {title} {\bibinfo {title} {{Augmented Extraction Efficiency of a Hot D
  Exciton in MoS${}_2$ via Intervalley Scattering}},\ }\href
  {https://doi.org/10.1021/acs.nanolett.4c01837} {\bibfield  {journal}
  {\bibinfo  {journal} {Nano Letters}\ }\textbf {\bibinfo {volume} {24}},\
  \bibinfo {pages} {11163} (\bibinfo {year} {2024})},\ \bibinfo {note} {pMID:
  39225119},\ \Eprint
  {https://arxiv.org/abs/https://doi.org/10.1021/acs.nanolett.4c01837}
  {https://doi.org/10.1021/acs.nanolett.4c01837} \BibitemShut {NoStop}%
\bibitem [{\citenamefont {Goswami}\ \emph {et~al.}(2021)\citenamefont
  {Goswami}, \citenamefont {Bhatt}, \citenamefont {Babu}, \citenamefont {Kaur},
  \citenamefont {Ghorai},\ and\ \citenamefont {Ghosh}}]{Goswami2021}%
  \BibitemOpen
  \bibfield  {author} {\bibinfo {author} {\bibfnamefont {T.}~\bibnamefont
  {Goswami}}, \bibinfo {author} {\bibfnamefont {H.}~\bibnamefont {Bhatt}},
  \bibinfo {author} {\bibfnamefont {K.~J.}\ \bibnamefont {Babu}}, \bibinfo
  {author} {\bibfnamefont {G.}~\bibnamefont {Kaur}}, \bibinfo {author}
  {\bibfnamefont {N.}~\bibnamefont {Ghorai}},\ and\ \bibinfo {author}
  {\bibfnamefont {H.~N.}\ \bibnamefont {Ghosh}},\ }\bibfield  {title} {\bibinfo
  {title} {{Ultrafast Insights into High Energy (C and D) Excitons in Few Layer
  WS${}_2$}},\ }\href {https://doi.org/10.1021/acs.jpclett.1c01627} {\bibfield
  {journal} {\bibinfo  {journal} {Journal of Physical Chemistry Letters}\
  }\textbf {\bibinfo {volume} {12}},\ \bibinfo {pages} {6526 – 6534}
  (\bibinfo {year} {2021})}\BibitemShut {NoStop}%
\bibitem [{\citenamefont {Chen}\ \emph {et~al.}(2018)\citenamefont {Chen},
  \citenamefont {Wang}, \citenamefont {Wang}, \citenamefont {Wang},
  \citenamefont {Yue}, \citenamefont {Wang}, \citenamefont {Wang},
  \citenamefont {Wee}, \citenamefont {Qiu},\ and\ \citenamefont
  {Sun}}]{chen2018investigating}%
  \BibitemOpen
  \bibfield  {author} {\bibinfo {author} {\bibfnamefont {X.}~\bibnamefont
  {Chen}}, \bibinfo {author} {\bibfnamefont {Z.}~\bibnamefont {Wang}}, \bibinfo
  {author} {\bibfnamefont {L.}~\bibnamefont {Wang}}, \bibinfo {author}
  {\bibfnamefont {H.-Y.}\ \bibnamefont {Wang}}, \bibinfo {author}
  {\bibfnamefont {Y.-Y.}\ \bibnamefont {Yue}}, \bibinfo {author} {\bibfnamefont
  {H.}~\bibnamefont {Wang}}, \bibinfo {author} {\bibfnamefont {X.-P.}\
  \bibnamefont {Wang}}, \bibinfo {author} {\bibfnamefont {A.~T.~S.}\
  \bibnamefont {Wee}}, \bibinfo {author} {\bibfnamefont {C.-W.}\ \bibnamefont
  {Qiu}},\ and\ \bibinfo {author} {\bibfnamefont {H.-B.}\ \bibnamefont {Sun}},\
  }\bibfield  {title} {\bibinfo {title} {{Investigating the dynamics of
  excitons in monolayer WSe${}_2$ before and after organic super acid
  treatment}},\ }\href {https://doi.org/10.1039/C8NR00774H} {\bibfield
  {journal} {\bibinfo  {journal} {Nanoscale}\ }\textbf {\bibinfo {volume}
  {10}},\ \bibinfo {pages} {9346} (\bibinfo {year} {2018})}\BibitemShut
  {NoStop}%
\bibitem [{\citenamefont {Chernikov}\ \emph {et~al.}(2015)\citenamefont
  {Chernikov}, \citenamefont {Ruppert}, \citenamefont {Hill}, \citenamefont
  {Rigosi},\ and\ \citenamefont {Heinz}}]{chernikov2015population}%
  \BibitemOpen
  \bibfield  {author} {\bibinfo {author} {\bibfnamefont {A.}~\bibnamefont
  {Chernikov}}, \bibinfo {author} {\bibfnamefont {C.}~\bibnamefont {Ruppert}},
  \bibinfo {author} {\bibfnamefont {H.~M.}\ \bibnamefont {Hill}}, \bibinfo
  {author} {\bibfnamefont {A.~F.}\ \bibnamefont {Rigosi}},\ and\ \bibinfo
  {author} {\bibfnamefont {T.~F.}\ \bibnamefont {Heinz}},\ }\bibfield  {title}
  {\bibinfo {title} {{Population inversion and giant bandgap renormalization in
  atomically thin WS${}_2$ layers}},\ }\href
  {https://doi.org/10.1038/nphoton.2015.104} {\bibfield  {journal} {\bibinfo
  {journal} {Nature Photonics}\ }\textbf {\bibinfo {volume} {9}},\ \bibinfo
  {pages} {466} (\bibinfo {year} {2015})}\BibitemShut {NoStop}%
\bibitem [{\citenamefont {Steinhoff}\ \emph {et~al.}(2017)\citenamefont
  {Steinhoff}, \citenamefont {Florian}, \citenamefont {R{\"o}sner},
  \citenamefont {Sch{\"o}nhoff}, \citenamefont {Wehling},\ and\ \citenamefont
  {Jahnke}}]{steinhoff2017exciton}%
  \BibitemOpen
  \bibfield  {author} {\bibinfo {author} {\bibfnamefont {A.}~\bibnamefont
  {Steinhoff}}, \bibinfo {author} {\bibfnamefont {M.}~\bibnamefont {Florian}},
  \bibinfo {author} {\bibfnamefont {M.}~\bibnamefont {R{\"o}sner}}, \bibinfo
  {author} {\bibfnamefont {G.}~\bibnamefont {Sch{\"o}nhoff}}, \bibinfo {author}
  {\bibfnamefont {T.~O.}\ \bibnamefont {Wehling}},\ and\ \bibinfo {author}
  {\bibfnamefont {F.}~\bibnamefont {Jahnke}},\ }\bibfield  {title} {\bibinfo
  {title} {Exciton fission in monolayer transition metal dichalcogenide
  semiconductors},\ }\href {https://doi.org/10.1038/s41467-017-01298-6}
  {\bibfield  {journal} {\bibinfo  {journal} {Nature Communications}\ }\textbf
  {\bibinfo {volume} {8}},\ \bibinfo {pages} {1166} (\bibinfo {year}
  {2017})}\BibitemShut {NoStop}%
\bibitem [{\citenamefont {Calati}\ \emph {et~al.}(2023)\citenamefont {Calati},
  \citenamefont {Li}, \citenamefont {Zhu},\ and\ \citenamefont
  {St\"ahler}}]{Calati2023}%
  \BibitemOpen
  \bibfield  {author} {\bibinfo {author} {\bibfnamefont {S.}~\bibnamefont
  {Calati}}, \bibinfo {author} {\bibfnamefont {Q.}~\bibnamefont {Li}}, \bibinfo
  {author} {\bibfnamefont {X.}~\bibnamefont {Zhu}},\ and\ \bibinfo {author}
  {\bibfnamefont {J.}~\bibnamefont {St\"ahler}},\ }\bibfield  {title} {\bibinfo
  {title} {{Dynamic screening of quasiparticles in ${\mathrm{WS}}_{2}$
  monolayers}},\ }\href@noop {} {\bibfield  {journal} {\bibinfo  {journal}
  {Phys. Rev. B}\ }\textbf {\bibinfo {volume} {107}},\ \bibinfo {pages}
  {115404} (\bibinfo {year} {2023})}\BibitemShut {NoStop}%
\bibitem [{\citenamefont {Lin}\ \emph {et~al.}(2021)\citenamefont {Lin},
  \citenamefont {Ong}, \citenamefont {Bange}, \citenamefont {Faria~Junior},
  \citenamefont {Peng}, \citenamefont {Ziegler}, \citenamefont {Zipfel},
  \citenamefont {B{\"a}uml}, \citenamefont {Paradiso}, \citenamefont
  {Watanabe}, \citenamefont {Taniguchi}, \citenamefont {Strunk}, \citenamefont
  {Monserrat}, \citenamefont {Fabian}, \citenamefont {Chernikov}, \citenamefont
  {Qiu}, \citenamefont {Louie},\ and\ \citenamefont {Lupton}}]{lin2021}%
  \BibitemOpen
  \bibfield  {author} {\bibinfo {author} {\bibfnamefont {K.-Q.}\ \bibnamefont
  {Lin}}, \bibinfo {author} {\bibfnamefont {C.~S.}\ \bibnamefont {Ong}},
  \bibinfo {author} {\bibfnamefont {S.}~\bibnamefont {Bange}}, \bibinfo
  {author} {\bibfnamefont {P.~E.}\ \bibnamefont {Faria~Junior}}, \bibinfo
  {author} {\bibfnamefont {B.}~\bibnamefont {Peng}}, \bibinfo {author}
  {\bibfnamefont {J.~D.}\ \bibnamefont {Ziegler}}, \bibinfo {author}
  {\bibfnamefont {J.}~\bibnamefont {Zipfel}}, \bibinfo {author} {\bibfnamefont
  {C.}~\bibnamefont {B{\"a}uml}}, \bibinfo {author} {\bibfnamefont
  {N.}~\bibnamefont {Paradiso}}, \bibinfo {author} {\bibfnamefont
  {K.}~\bibnamefont {Watanabe}}, \bibinfo {author} {\bibfnamefont
  {T.}~\bibnamefont {Taniguchi}}, \bibinfo {author} {\bibfnamefont
  {C.}~\bibnamefont {Strunk}}, \bibinfo {author} {\bibfnamefont
  {B.}~\bibnamefont {Monserrat}}, \bibinfo {author} {\bibfnamefont
  {J.}~\bibnamefont {Fabian}}, \bibinfo {author} {\bibfnamefont
  {A.}~\bibnamefont {Chernikov}}, \bibinfo {author} {\bibfnamefont {D.~Y.}\
  \bibnamefont {Qiu}}, \bibinfo {author} {\bibfnamefont {S.~G.}\ \bibnamefont
  {Louie}},\ and\ \bibinfo {author} {\bibfnamefont {J.~M.}\ \bibnamefont
  {Lupton}},\ }\bibfield  {title} {\bibinfo {title} {{Narrow-band high-lying
  excitons with negative-mass electrons in monolayer WSe${}_2$}},\ }\href
  {https://doi.org/10.1038/s41467-021-25499-2} {\bibfield  {journal} {\bibinfo
  {journal} {Nature Communications}\ }\textbf {\bibinfo {volume} {12}},\
  \bibinfo {pages} {5500} (\bibinfo {year} {2021})}\BibitemShut {NoStop}%
\bibitem [{sup()}]{supplementary_citations}%
  \BibitemOpen
  \href@noop {} {}\bibinfo {howpublished} {Supplementary Material, which
  includes Refs.~\cite{manzoni2016, giannozzi2009quantum,
  giannozzi2017advanced, giannozzi2020quantum, perdew1996generalized,
  van2018pseudodojo, yun2012thickness, monkhorst1976special, reining2018gw,
  onida2002electronic, strinati1988application, marini_yambo_2009,
  sangalli2019many, marsili2021spinorial, stan2009levels, PPA_yambo, RPA,
  guandalini2022efficient, bruneval2008accurate}, with extended methods and
  additional results.}\BibitemShut {Stop}%
\bibitem [{\citenamefont {O’Donnell}\ and\ \citenamefont
  {Chen}(1991)}]{odonnell1991}%
  \BibitemOpen
  \bibfield  {author} {\bibinfo {author} {\bibfnamefont {K.~P.}\ \bibnamefont
  {O’Donnell}}\ and\ \bibinfo {author} {\bibfnamefont {X.}~\bibnamefont
  {Chen}},\ }\bibfield  {title} {\bibinfo {title} {Temperature dependence of
  semiconductor band gaps},\ }\href {https://doi.org/10.1063/1.104723}
  {\bibfield  {journal} {\bibinfo  {journal} {Applied Physics Letters}\
  }\textbf {\bibinfo {volume} {58}},\ \bibinfo {pages} {2924} (\bibinfo {year}
  {1991})},\ \Eprint
  {https://arxiv.org/abs/https://pubs.aip.org/aip/apl/article-pdf/58/25/2924/18483026/2924\_1\_online.pdf}
  {https://pubs.aip.org/aip/apl/article-pdf/58/25/2924/18483026/2924\_1\_online.pdf}
  \BibitemShut {NoStop}%
\bibitem [{\citenamefont {Nguyen}\ \emph {et~al.}(2024)\citenamefont {Nguyen},
  \citenamefont {Le}, \citenamefont {Kim}, \citenamefont {Kim}, \citenamefont
  {Diware}, \citenamefont {Kim},\ and\ \citenamefont {Kim}}]{nguyen2024}%
  \BibitemOpen
  \bibfield  {author} {\bibinfo {author} {\bibfnamefont {X.~A.}\ \bibnamefont
  {Nguyen}}, \bibinfo {author} {\bibfnamefont {L.~V.}\ \bibnamefont {Le}},
  \bibinfo {author} {\bibfnamefont {S.~H.}\ \bibnamefont {Kim}}, \bibinfo
  {author} {\bibfnamefont {Y.~D.}\ \bibnamefont {Kim}}, \bibinfo {author}
  {\bibfnamefont {M.~S.}\ \bibnamefont {Diware}}, \bibinfo {author}
  {\bibfnamefont {T.~J.}\ \bibnamefont {Kim}},\ and\ \bibinfo {author}
  {\bibfnamefont {Y.~D.}\ \bibnamefont {Kim}},\ }\bibfield  {title} {\bibinfo
  {title} {{Temperature dependence of the dielectric function and critical
  points of monolayer WSe${}_2$}},\ }\href
  {https://doi.org/10.1038/s41598-024-64303-1} {\bibfield  {journal} {\bibinfo
  {journal} {Scientific Reports}\ }\textbf {\bibinfo {volume} {14}},\ \bibinfo
  {pages} {13486} (\bibinfo {year} {2024})}\BibitemShut {NoStop}%
\bibitem [{\citenamefont {Dogadov}\ \emph
  {et~al.}(2026{\natexlab{b}})\citenamefont {Dogadov}, \citenamefont
  {Cervantes-Villanueva}, \citenamefont {Molina-Sánchez}, \citenamefont
  {Sangalli},\ and\ \citenamefont {Dal~Conte}}]{dogadov_2026_19815236}%
  \BibitemOpen
  \bibfield  {author} {\bibinfo {author} {\bibfnamefont {O.}~\bibnamefont
  {Dogadov}}, \bibinfo {author} {\bibfnamefont {J.}~\bibnamefont
  {Cervantes-Villanueva}}, \bibinfo {author} {\bibfnamefont {A.}~\bibnamefont
  {Molina-Sánchez}}, \bibinfo {author} {\bibfnamefont {D.}~\bibnamefont
  {Sangalli}},\ and\ \bibinfo {author} {\bibfnamefont {S.}~\bibnamefont
  {Dal~Conte}},\ }\bibfield  {title} {\bibinfo {title} {{Data set of the
  publication "Nonequilibrium dynamics of high energy transitions in monolayer
  WSe${}_2$" }},\ }\href {https://doi.org/10.5281/zenodo.19815236}
  {10.5281/zenodo.19815236} (\bibinfo {year} {2026}{\natexlab{b}})\BibitemShut
  {NoStop}%
\bibitem [{\citenamefont {Malic}\ \emph {et~al.}(2018)\citenamefont {Malic},
  \citenamefont {Selig}, \citenamefont {Feierabend}, \citenamefont {Brem},
  \citenamefont {Christiansen}, \citenamefont {Wendler}, \citenamefont
  {Knorr},\ and\ \citenamefont {Bergh\"auser}}]{Malic2018}%
  \BibitemOpen
  \bibfield  {author} {\bibinfo {author} {\bibfnamefont {E.}~\bibnamefont
  {Malic}}, \bibinfo {author} {\bibfnamefont {M.}~\bibnamefont {Selig}},
  \bibinfo {author} {\bibfnamefont {M.}~\bibnamefont {Feierabend}}, \bibinfo
  {author} {\bibfnamefont {S.}~\bibnamefont {Brem}}, \bibinfo {author}
  {\bibfnamefont {D.}~\bibnamefont {Christiansen}}, \bibinfo {author}
  {\bibfnamefont {F.}~\bibnamefont {Wendler}}, \bibinfo {author} {\bibfnamefont
  {A.}~\bibnamefont {Knorr}},\ and\ \bibinfo {author} {\bibfnamefont
  {G.}~\bibnamefont {Bergh\"auser}},\ }\bibfield  {title} {\bibinfo {title}
  {Dark excitons in transition metal dichalcogenides},\ }\href
  {https://doi.org/10.1103/PhysRevMaterials.2.014002} {\bibfield  {journal}
  {\bibinfo  {journal} {Phys. Rev. Mater.}\ }\textbf {\bibinfo {volume} {2}},\
  \bibinfo {pages} {014002} (\bibinfo {year} {2018})}\BibitemShut {NoStop}%
\bibitem [{\citenamefont {Selig}\ \emph {et~al.}(2018)\citenamefont {Selig},
  \citenamefont {Berghäuser}, \citenamefont {Richter}, \citenamefont
  {Bratschitsch}, \citenamefont {Knorr},\ and\ \citenamefont
  {Malic}}]{Selig2018}%
  \BibitemOpen
  \bibfield  {author} {\bibinfo {author} {\bibfnamefont {M.}~\bibnamefont
  {Selig}}, \bibinfo {author} {\bibfnamefont {G.}~\bibnamefont {Berghäuser}},
  \bibinfo {author} {\bibfnamefont {M.}~\bibnamefont {Richter}}, \bibinfo
  {author} {\bibfnamefont {R.}~\bibnamefont {Bratschitsch}}, \bibinfo {author}
  {\bibfnamefont {A.}~\bibnamefont {Knorr}},\ and\ \bibinfo {author}
  {\bibfnamefont {E.}~\bibnamefont {Malic}},\ }\bibfield  {title} {\bibinfo
  {title} {Dark and bright exciton formation, thermalization, and
  photoluminescence in monolayer transition metal dichalcogenides},\ }\href
  {https://doi.org/10.1088/2053-1583/aabea3} {\bibfield  {journal} {\bibinfo
  {journal} {2D Materials}\ }\textbf {\bibinfo {volume} {5}},\ \bibinfo {pages}
  {035017} (\bibinfo {year} {2018})}\BibitemShut {NoStop}%
\bibitem [{\citenamefont {Madéo}\ \emph {et~al.}(2020)\citenamefont {Madéo},
  \citenamefont {Man}, \citenamefont {Sahoo}, \citenamefont {Campbell},
  \citenamefont {Pareek}, \citenamefont {Wong}, \citenamefont {Al-Mahboob},
  \citenamefont {Chan}, \citenamefont {Karmakar}, \citenamefont {Mariserla},
  \citenamefont {Li}, \citenamefont {Heinz}, \citenamefont {Cao},\ and\
  \citenamefont {Dani}}]{Madeo2020}%
  \BibitemOpen
  \bibfield  {author} {\bibinfo {author} {\bibfnamefont {J.}~\bibnamefont
  {Madéo}}, \bibinfo {author} {\bibfnamefont {M.~K.~L.}\ \bibnamefont {Man}},
  \bibinfo {author} {\bibfnamefont {C.}~\bibnamefont {Sahoo}}, \bibinfo
  {author} {\bibfnamefont {M.}~\bibnamefont {Campbell}}, \bibinfo {author}
  {\bibfnamefont {V.}~\bibnamefont {Pareek}}, \bibinfo {author} {\bibfnamefont
  {E.~L.}\ \bibnamefont {Wong}}, \bibinfo {author} {\bibfnamefont
  {A.}~\bibnamefont {Al-Mahboob}}, \bibinfo {author} {\bibfnamefont {N.~S.}\
  \bibnamefont {Chan}}, \bibinfo {author} {\bibfnamefont {A.}~\bibnamefont
  {Karmakar}}, \bibinfo {author} {\bibfnamefont {B.~M.~K.}\ \bibnamefont
  {Mariserla}}, \bibinfo {author} {\bibfnamefont {X.}~\bibnamefont {Li}},
  \bibinfo {author} {\bibfnamefont {T.~F.}\ \bibnamefont {Heinz}}, \bibinfo
  {author} {\bibfnamefont {T.}~\bibnamefont {Cao}},\ and\ \bibinfo {author}
  {\bibfnamefont {K.~M.}\ \bibnamefont {Dani}},\ }\bibfield  {title} {\bibinfo
  {title} {Directly visualizing the momentum-forbidden dark excitons and their
  dynamics in atomically thin semiconductors},\ }\href
  {https://doi.org/10.1126/science.aba1029} {\bibfield  {journal} {\bibinfo
  {journal} {Science}\ }\textbf {\bibinfo {volume} {370}},\ \bibinfo {pages}
  {1199} (\bibinfo {year} {2020})}\BibitemShut {NoStop}%
\bibitem [{\citenamefont {Bange}\ \emph {et~al.}(2023)\citenamefont {Bange},
  \citenamefont {Werner}, \citenamefont {Schmitt}, \citenamefont {Bennecke},
  \citenamefont {Meneghini}, \citenamefont {AlMutairi}, \citenamefont
  {Merboldt}, \citenamefont {Watanabe}, \citenamefont {Taniguchi},
  \citenamefont {Steil}, \citenamefont {Steil}, \citenamefont {Weitz},
  \citenamefont {Hofmann}, \citenamefont {Jansen}, \citenamefont {Brem},
  \citenamefont {Malic}, \citenamefont {Reutzel},\ and\ \citenamefont
  {Mathias}}]{Bange2023}%
  \BibitemOpen
  \bibfield  {author} {\bibinfo {author} {\bibfnamefont {J.~P.}\ \bibnamefont
  {Bange}}, \bibinfo {author} {\bibfnamefont {P.}~\bibnamefont {Werner}},
  \bibinfo {author} {\bibfnamefont {D.}~\bibnamefont {Schmitt}}, \bibinfo
  {author} {\bibfnamefont {W.}~\bibnamefont {Bennecke}}, \bibinfo {author}
  {\bibfnamefont {G.}~\bibnamefont {Meneghini}}, \bibinfo {author}
  {\bibfnamefont {A.}~\bibnamefont {AlMutairi}}, \bibinfo {author}
  {\bibfnamefont {M.}~\bibnamefont {Merboldt}}, \bibinfo {author}
  {\bibfnamefont {K.}~\bibnamefont {Watanabe}}, \bibinfo {author}
  {\bibfnamefont {T.}~\bibnamefont {Taniguchi}}, \bibinfo {author}
  {\bibfnamefont {S.}~\bibnamefont {Steil}}, \bibinfo {author} {\bibfnamefont
  {D.}~\bibnamefont {Steil}}, \bibinfo {author} {\bibfnamefont {R.~T.}\
  \bibnamefont {Weitz}}, \bibinfo {author} {\bibfnamefont {S.}~\bibnamefont
  {Hofmann}}, \bibinfo {author} {\bibfnamefont {G.~S.~M.}\ \bibnamefont
  {Jansen}}, \bibinfo {author} {\bibfnamefont {S.}~\bibnamefont {Brem}},
  \bibinfo {author} {\bibfnamefont {E.}~\bibnamefont {Malic}}, \bibinfo
  {author} {\bibfnamefont {M.}~\bibnamefont {Reutzel}},\ and\ \bibinfo {author}
  {\bibfnamefont {S.}~\bibnamefont {Mathias}},\ }\bibfield  {title} {\bibinfo
  {title} {{Ultrafast dynamics of bright and dark excitons in monolayer
  WSe${}_2$ and heterobilayer WSe${}_2$/MoS${}_2$}},\ }\href
  {https://doi.org/10.1088/2053-1583/ace067} {\bibfield  {journal} {\bibinfo
  {journal} {2D Materials}\ }\textbf {\bibinfo {volume} {10}},\ \bibinfo
  {pages} {035039} (\bibinfo {year} {2023})}\BibitemShut {NoStop}%
\bibitem [{\citenamefont {Sangalli}\ \emph {et~al.}(2016)\citenamefont
  {Sangalli}, \citenamefont {Dal~Conte}, \citenamefont {Manzoni}, \citenamefont
  {Cerullo},\ and\ \citenamefont {Marini}}]{Sangalli2016}%
  \BibitemOpen
  \bibfield  {author} {\bibinfo {author} {\bibfnamefont {D.}~\bibnamefont
  {Sangalli}}, \bibinfo {author} {\bibfnamefont {S.}~\bibnamefont {Dal~Conte}},
  \bibinfo {author} {\bibfnamefont {C.}~\bibnamefont {Manzoni}}, \bibinfo
  {author} {\bibfnamefont {G.}~\bibnamefont {Cerullo}},\ and\ \bibinfo {author}
  {\bibfnamefont {A.}~\bibnamefont {Marini}},\ }\bibfield  {title} {\bibinfo
  {title} {Nonequilibrium optical properties in semiconductors from first
  principles: A combined theoretical and experimental study of bulk silicon},\
  }\href {https://doi.org/10.1103/PhysRevB.93.195205} {\bibfield  {journal}
  {\bibinfo  {journal} {Phys. Rev. B}\ }\textbf {\bibinfo {volume} {93}},\
  \bibinfo {pages} {195205} (\bibinfo {year} {2016})}\BibitemShut {NoStop}%
\bibitem [{\citenamefont {Wang}\ and\ \citenamefont
  {Zhang}(2024)}]{wang2024experimental}%
  \BibitemOpen
  \bibfield  {author} {\bibinfo {author} {\bibfnamefont {Y.}~\bibnamefont
  {Wang}}\ and\ \bibinfo {author} {\bibfnamefont {X.}~\bibnamefont {Zhang}},\
  }\bibfield  {title} {\bibinfo {title} {{Experimental and Theoretical
  Investigations of Direct and Indirect Band Gaps of WSe${}_2$}},\ }\bibfield
  {journal} {\bibinfo  {journal} {Micromachines}\ }\textbf {\bibinfo {volume}
  {15}},\ \href {https://doi.org/10.3390/mi15060761} {10.3390/mi15060761}
  (\bibinfo {year} {2024})\BibitemShut {NoStop}%
\bibitem [{\citenamefont {Tonndorf}\ \emph {et~al.}(2013)\citenamefont
  {Tonndorf}, \citenamefont {Schmidt}, \citenamefont {B{\"o}ttger},
  \citenamefont {Zhang}, \citenamefont {B{\"o}rner}, \citenamefont {Liebig},
  \citenamefont {Albrecht}, \citenamefont {Kloc}, \citenamefont {Gordan},
  \citenamefont {Zahn} \emph {et~al.}}]{tonndorf2013photoluminescence}%
  \BibitemOpen
  \bibfield  {author} {\bibinfo {author} {\bibfnamefont {P.}~\bibnamefont
  {Tonndorf}}, \bibinfo {author} {\bibfnamefont {R.}~\bibnamefont {Schmidt}},
  \bibinfo {author} {\bibfnamefont {P.}~\bibnamefont {B{\"o}ttger}}, \bibinfo
  {author} {\bibfnamefont {X.}~\bibnamefont {Zhang}}, \bibinfo {author}
  {\bibfnamefont {J.}~\bibnamefont {B{\"o}rner}}, \bibinfo {author}
  {\bibfnamefont {A.}~\bibnamefont {Liebig}}, \bibinfo {author} {\bibfnamefont
  {M.}~\bibnamefont {Albrecht}}, \bibinfo {author} {\bibfnamefont
  {C.}~\bibnamefont {Kloc}}, \bibinfo {author} {\bibfnamefont {O.}~\bibnamefont
  {Gordan}}, \bibinfo {author} {\bibfnamefont {D.~R.}\ \bibnamefont {Zahn}},
  \emph {et~al.},\ }\bibfield  {title} {\bibinfo {title} {{Photoluminescence
  emission and Raman response of monolayer MoS${}_2$, MoSe${}_2$, and
  WSe${}_2$}},\ }\href@noop {} {\bibfield  {journal} {\bibinfo  {journal}
  {Optics express}\ }\textbf {\bibinfo {volume} {21}},\ \bibinfo {pages} {4908}
  (\bibinfo {year} {2013})}\BibitemShut {NoStop}%
\bibitem [{\citenamefont {Hsu}\ \emph {et~al.}(2017)\citenamefont {Hsu},
  \citenamefont {Lu}, \citenamefont {Wang}, \citenamefont {Huang},
  \citenamefont {Li}, \citenamefont {Chang}, \citenamefont {Chou},
  \citenamefont {Juang}, \citenamefont {Jeng}, \citenamefont {Li} \emph
  {et~al.}}]{hsu2017evidence}%
  \BibitemOpen
  \bibfield  {author} {\bibinfo {author} {\bibfnamefont {W.-T.}\ \bibnamefont
  {Hsu}}, \bibinfo {author} {\bibfnamefont {L.-S.}\ \bibnamefont {Lu}},
  \bibinfo {author} {\bibfnamefont {D.}~\bibnamefont {Wang}}, \bibinfo {author}
  {\bibfnamefont {J.-K.}\ \bibnamefont {Huang}}, \bibinfo {author}
  {\bibfnamefont {M.-Y.}\ \bibnamefont {Li}}, \bibinfo {author} {\bibfnamefont
  {T.-R.}\ \bibnamefont {Chang}}, \bibinfo {author} {\bibfnamefont {Y.-C.}\
  \bibnamefont {Chou}}, \bibinfo {author} {\bibfnamefont {Z.-Y.}\ \bibnamefont
  {Juang}}, \bibinfo {author} {\bibfnamefont {H.-T.}\ \bibnamefont {Jeng}},
  \bibinfo {author} {\bibfnamefont {L.-J.}\ \bibnamefont {Li}}, \emph
  {et~al.},\ }\bibfield  {title} {\bibinfo {title} {{Evidence of indirect gap
  in monolayer WSe${}_2$}},\ }\href@noop {} {\bibfield  {journal} {\bibinfo
  {journal} {Nature communications}\ }\textbf {\bibinfo {volume} {8}},\
  \bibinfo {pages} {929} (\bibinfo {year} {2017})}\BibitemShut {NoStop}%
\bibitem [{\citenamefont {Erben}\ \emph {et~al.}(2018)\citenamefont {Erben},
  \citenamefont {Steinhoff}, \citenamefont {Gies}, \citenamefont {Sch\"onhoff},
  \citenamefont {Wehling},\ and\ \citenamefont {Jahnke}}]{erben2018excitation}%
  \BibitemOpen
  \bibfield  {author} {\bibinfo {author} {\bibfnamefont {D.}~\bibnamefont
  {Erben}}, \bibinfo {author} {\bibfnamefont {A.}~\bibnamefont {Steinhoff}},
  \bibinfo {author} {\bibfnamefont {C.}~\bibnamefont {Gies}}, \bibinfo {author}
  {\bibfnamefont {G.}~\bibnamefont {Sch\"onhoff}}, \bibinfo {author}
  {\bibfnamefont {T.~O.}\ \bibnamefont {Wehling}},\ and\ \bibinfo {author}
  {\bibfnamefont {F.}~\bibnamefont {Jahnke}},\ }\bibfield  {title} {\bibinfo
  {title} {Excitation-induced transition to indirect band gaps in atomically
  thin transition-metal dichalcogenide semiconductors},\ }\href
  {https://doi.org/10.1103/PhysRevB.98.035434} {\bibfield  {journal} {\bibinfo
  {journal} {Phys. Rev. B}\ }\textbf {\bibinfo {volume} {98}},\ \bibinfo
  {pages} {035434} (\bibinfo {year} {2018})}\BibitemShut {NoStop}%
\bibitem [{\citenamefont {Muoi}\ \emph {et~al.}(2019)\citenamefont {Muoi},
  \citenamefont {Hieu}, \citenamefont {Phung}, \citenamefont {Phuc},
  \citenamefont {Amin}, \citenamefont {Hoi}, \citenamefont {Hieu},
  \citenamefont {Nhan}, \citenamefont {Nguyen},\ and\ \citenamefont
  {Le}}]{muoi2019electronic}%
  \BibitemOpen
  \bibfield  {author} {\bibinfo {author} {\bibfnamefont {D.}~\bibnamefont
  {Muoi}}, \bibinfo {author} {\bibfnamefont {N.~N.}\ \bibnamefont {Hieu}},
  \bibinfo {author} {\bibfnamefont {H.~T.}\ \bibnamefont {Phung}}, \bibinfo
  {author} {\bibfnamefont {H.~V.}\ \bibnamefont {Phuc}}, \bibinfo {author}
  {\bibfnamefont {B.}~\bibnamefont {Amin}}, \bibinfo {author} {\bibfnamefont
  {B.~D.}\ \bibnamefont {Hoi}}, \bibinfo {author} {\bibfnamefont {N.~V.}\
  \bibnamefont {Hieu}}, \bibinfo {author} {\bibfnamefont {L.~C.}\ \bibnamefont
  {Nhan}}, \bibinfo {author} {\bibfnamefont {C.~V.}\ \bibnamefont {Nguyen}},\
  and\ \bibinfo {author} {\bibfnamefont {P.}~\bibnamefont {Le}},\ }\bibfield
  {title} {\bibinfo {title} {{Electronic properties of WS${}_2$ and WSe${}_2$
  monolayers with biaxial strain: A first-principles study}},\ }\href
  {https://doi.org/https://doi.org/10.1016/j.chemphys.2018.12.004} {\bibfield
  {journal} {\bibinfo  {journal} {Chemical Physics}\ }\textbf {\bibinfo
  {volume} {519}},\ \bibinfo {pages} {69} (\bibinfo {year} {2019})}\BibitemShut
  {NoStop}%
\bibitem [{\citenamefont {Steinhoff}\ \emph {et~al.}(2015)\citenamefont
  {Steinhoff}, \citenamefont {Kim}, \citenamefont {Jahnke}, \citenamefont
  {R{\"o}sner}, \citenamefont {Kim}, \citenamefont {Lee}, \citenamefont {Han},
  \citenamefont {Jeong}, \citenamefont {Wehling},\ and\ \citenamefont
  {Gies}}]{steinhoff2015efficient}%
  \BibitemOpen
  \bibfield  {author} {\bibinfo {author} {\bibfnamefont {A.}~\bibnamefont
  {Steinhoff}}, \bibinfo {author} {\bibfnamefont {J.-H.}\ \bibnamefont {Kim}},
  \bibinfo {author} {\bibfnamefont {F.}~\bibnamefont {Jahnke}}, \bibinfo
  {author} {\bibfnamefont {M.}~\bibnamefont {R{\"o}sner}}, \bibinfo {author}
  {\bibfnamefont {D.-S.}\ \bibnamefont {Kim}}, \bibinfo {author} {\bibfnamefont
  {C.}~\bibnamefont {Lee}}, \bibinfo {author} {\bibfnamefont {G.~H.}\
  \bibnamefont {Han}}, \bibinfo {author} {\bibfnamefont {M.~S.}\ \bibnamefont
  {Jeong}}, \bibinfo {author} {\bibfnamefont {T.~O.}\ \bibnamefont {Wehling}},\
  and\ \bibinfo {author} {\bibfnamefont {C.}~\bibnamefont {Gies}},\ }\bibfield
  {title} {\bibinfo {title} {{Efficient Excitonic Photoluminescence in Direct
  and Indirect Band Gap Monolayer MoS${}_2$}},\ }\href
  {https://doi.org/10.1021/acs.nanolett.5b02719} {\bibfield  {journal}
  {\bibinfo  {journal} {Nano Letters}\ }\textbf {\bibinfo {volume} {15}},\
  \bibinfo {pages} {6841} (\bibinfo {year} {2015})},\ \bibinfo {note} {pMID:
  26322814},\ \Eprint
  {https://arxiv.org/abs/https://doi.org/10.1021/acs.nanolett.5b02719}
  {https://doi.org/10.1021/acs.nanolett.5b02719} \BibitemShut {NoStop}%
\bibitem [{\citenamefont {Manzoni}\ and\ \citenamefont
  {Cerullo}(2016)}]{manzoni2016}%
  \BibitemOpen
  \bibfield  {author} {\bibinfo {author} {\bibfnamefont {C.}~\bibnamefont
  {Manzoni}}\ and\ \bibinfo {author} {\bibfnamefont {G.}~\bibnamefont
  {Cerullo}},\ }\bibfield  {title} {\bibinfo {title} {Design criteria for
  ultrafast optical parametric amplifiers},\ }\href
  {https://doi.org/10.1088/2040-8978/18/10/103501} {\bibfield  {journal}
  {\bibinfo  {journal} {Journal of Optics}\ }\textbf {\bibinfo {volume} {18}},\
  \bibinfo {pages} {103501} (\bibinfo {year} {2016})}\BibitemShut {NoStop}%
\bibitem [{\citenamefont {Giannozzi}\ \emph {et~al.}(2009)\citenamefont
  {Giannozzi}, \citenamefont {Baroni}, \citenamefont {Bonini}, \citenamefont
  {Calandra}, \citenamefont {Car}, \citenamefont {Cavazzoni}, \citenamefont
  {Ceresoli}, \citenamefont {Chiarotti}, \citenamefont {Cococcioni},
  \citenamefont {Dabo}, \citenamefont {Corso}, \citenamefont {de~Gironcoli},
  \citenamefont {Fabris}, \citenamefont {Fratesi}, \citenamefont {Gebauer},
  \citenamefont {Gerstmann}, \citenamefont {Gougoussis}, \citenamefont
  {Kokalj}, \citenamefont {Lazzeri}, \citenamefont {Martin-Samos},
  \citenamefont {Marzari}, \citenamefont {Mauri}, \citenamefont {Mazzarello},
  \citenamefont {Paolini}, \citenamefont {Pasquarello}, \citenamefont
  {Paulatto}, \citenamefont {Sbraccia}, \citenamefont {Scandolo}, \citenamefont
  {Sclauzero}, \citenamefont {Seitsonen}, \citenamefont {Smogunov},
  \citenamefont {Umari},\ and\ \citenamefont
  {Wentzcovitch}}]{giannozzi2009quantum}%
  \BibitemOpen
  \bibfield  {author} {\bibinfo {author} {\bibfnamefont {P.}~\bibnamefont
  {Giannozzi}}, \bibinfo {author} {\bibfnamefont {S.}~\bibnamefont {Baroni}},
  \bibinfo {author} {\bibfnamefont {N.}~\bibnamefont {Bonini}}, \bibinfo
  {author} {\bibfnamefont {M.}~\bibnamefont {Calandra}}, \bibinfo {author}
  {\bibfnamefont {R.}~\bibnamefont {Car}}, \bibinfo {author} {\bibfnamefont
  {C.}~\bibnamefont {Cavazzoni}}, \bibinfo {author} {\bibfnamefont
  {D.}~\bibnamefont {Ceresoli}}, \bibinfo {author} {\bibfnamefont {G.~L.}\
  \bibnamefont {Chiarotti}}, \bibinfo {author} {\bibfnamefont {M.}~\bibnamefont
  {Cococcioni}}, \bibinfo {author} {\bibfnamefont {I.}~\bibnamefont {Dabo}},
  \bibinfo {author} {\bibfnamefont {A.~D.}\ \bibnamefont {Corso}}, \bibinfo
  {author} {\bibfnamefont {S.}~\bibnamefont {de~Gironcoli}}, \bibinfo {author}
  {\bibfnamefont {S.}~\bibnamefont {Fabris}}, \bibinfo {author} {\bibfnamefont
  {G.}~\bibnamefont {Fratesi}}, \bibinfo {author} {\bibfnamefont
  {R.}~\bibnamefont {Gebauer}}, \bibinfo {author} {\bibfnamefont
  {U.}~\bibnamefont {Gerstmann}}, \bibinfo {author} {\bibfnamefont
  {C.}~\bibnamefont {Gougoussis}}, \bibinfo {author} {\bibfnamefont
  {A.}~\bibnamefont {Kokalj}}, \bibinfo {author} {\bibfnamefont
  {M.}~\bibnamefont {Lazzeri}}, \bibinfo {author} {\bibfnamefont
  {L.}~\bibnamefont {Martin-Samos}}, \bibinfo {author} {\bibfnamefont
  {N.}~\bibnamefont {Marzari}}, \bibinfo {author} {\bibfnamefont
  {F.}~\bibnamefont {Mauri}}, \bibinfo {author} {\bibfnamefont
  {R.}~\bibnamefont {Mazzarello}}, \bibinfo {author} {\bibfnamefont
  {S.}~\bibnamefont {Paolini}}, \bibinfo {author} {\bibfnamefont
  {A.}~\bibnamefont {Pasquarello}}, \bibinfo {author} {\bibfnamefont
  {L.}~\bibnamefont {Paulatto}}, \bibinfo {author} {\bibfnamefont
  {C.}~\bibnamefont {Sbraccia}}, \bibinfo {author} {\bibfnamefont
  {S.}~\bibnamefont {Scandolo}}, \bibinfo {author} {\bibfnamefont
  {G.}~\bibnamefont {Sclauzero}}, \bibinfo {author} {\bibfnamefont {A.~P.}\
  \bibnamefont {Seitsonen}}, \bibinfo {author} {\bibfnamefont {A.}~\bibnamefont
  {Smogunov}}, \bibinfo {author} {\bibfnamefont {P.}~\bibnamefont {Umari}},\
  and\ \bibinfo {author} {\bibfnamefont {R.~M.}\ \bibnamefont {Wentzcovitch}},\
  }\bibfield  {title} {\bibinfo {title} {{QUANTUM ESPRESSO}: a modular and
  open-source software project for quantum simulations of materials},\ }\href
  {https://doi.org/10.1088/0953-8984/21/39/395502} {\bibfield  {journal}
  {\bibinfo  {journal} {Journal of Physics: Condensed Matter}\ }\textbf
  {\bibinfo {volume} {21}},\ \bibinfo {pages} {395502} (\bibinfo {year}
  {2009})}\BibitemShut {NoStop}%
\bibitem [{\citenamefont {Giannozzi}\ \emph {et~al.}(2017)\citenamefont
  {Giannozzi}, \citenamefont {Andreussi}, \citenamefont {Brumme}, \citenamefont
  {Bunau}, \citenamefont {Nardelli}, \citenamefont {Calandra}, \citenamefont
  {Car}, \citenamefont {Cavazzoni}, \citenamefont {Ceresoli}, \citenamefont
  {Cococcioni}, \citenamefont {Colonna}, \citenamefont {Carnimeo},
  \citenamefont {Corso}, \citenamefont {de~Gironcoli}, \citenamefont {Delugas},
  \citenamefont {DiStasio}, \citenamefont {Ferretti}, \citenamefont {Floris},
  \citenamefont {Fratesi}, \citenamefont {Fugallo}, \citenamefont {Gebauer},
  \citenamefont {Gerstmann}, \citenamefont {Giustino}, \citenamefont {Gorni},
  \citenamefont {Jia}, \citenamefont {Kawamura}, \citenamefont {Ko},
  \citenamefont {Kokalj}, \citenamefont {Küçükbenli}, \citenamefont
  {Lazzeri}, \citenamefont {Marsili}, \citenamefont {Marzari}, \citenamefont
  {Mauri}, \citenamefont {Nguyen}, \citenamefont {Nguyen}, \citenamefont {de-la
  Roza}, \citenamefont {Paulatto}, \citenamefont {Poncé}, \citenamefont
  {Rocca}, \citenamefont {Sabatini}, \citenamefont {Santra}, \citenamefont
  {Schlipf}, \citenamefont {Seitsonen}, \citenamefont {Smogunov}, \citenamefont
  {Timrov}, \citenamefont {Thonhauser}, \citenamefont {Umari}, \citenamefont
  {Vast}, \citenamefont {Wu},\ and\ \citenamefont
  {Baroni}}]{giannozzi2017advanced}%
  \BibitemOpen
  \bibfield  {author} {\bibinfo {author} {\bibfnamefont {P.}~\bibnamefont
  {Giannozzi}}, \bibinfo {author} {\bibfnamefont {O.}~\bibnamefont
  {Andreussi}}, \bibinfo {author} {\bibfnamefont {T.}~\bibnamefont {Brumme}},
  \bibinfo {author} {\bibfnamefont {O.}~\bibnamefont {Bunau}}, \bibinfo
  {author} {\bibfnamefont {M.~B.}\ \bibnamefont {Nardelli}}, \bibinfo {author}
  {\bibfnamefont {M.}~\bibnamefont {Calandra}}, \bibinfo {author}
  {\bibfnamefont {R.}~\bibnamefont {Car}}, \bibinfo {author} {\bibfnamefont
  {C.}~\bibnamefont {Cavazzoni}}, \bibinfo {author} {\bibfnamefont
  {D.}~\bibnamefont {Ceresoli}}, \bibinfo {author} {\bibfnamefont
  {M.}~\bibnamefont {Cococcioni}}, \bibinfo {author} {\bibfnamefont
  {N.}~\bibnamefont {Colonna}}, \bibinfo {author} {\bibfnamefont
  {I.}~\bibnamefont {Carnimeo}}, \bibinfo {author} {\bibfnamefont {A.~D.}\
  \bibnamefont {Corso}}, \bibinfo {author} {\bibfnamefont {S.}~\bibnamefont
  {de~Gironcoli}}, \bibinfo {author} {\bibfnamefont {P.}~\bibnamefont
  {Delugas}}, \bibinfo {author} {\bibfnamefont {R.~A.}\ \bibnamefont
  {DiStasio}}, \bibinfo {author} {\bibfnamefont {A.}~\bibnamefont {Ferretti}},
  \bibinfo {author} {\bibfnamefont {A.}~\bibnamefont {Floris}}, \bibinfo
  {author} {\bibfnamefont {G.}~\bibnamefont {Fratesi}}, \bibinfo {author}
  {\bibfnamefont {G.}~\bibnamefont {Fugallo}}, \bibinfo {author} {\bibfnamefont
  {R.}~\bibnamefont {Gebauer}}, \bibinfo {author} {\bibfnamefont
  {U.}~\bibnamefont {Gerstmann}}, \bibinfo {author} {\bibfnamefont
  {F.}~\bibnamefont {Giustino}}, \bibinfo {author} {\bibfnamefont
  {T.}~\bibnamefont {Gorni}}, \bibinfo {author} {\bibfnamefont
  {J.}~\bibnamefont {Jia}}, \bibinfo {author} {\bibfnamefont {M.}~\bibnamefont
  {Kawamura}}, \bibinfo {author} {\bibfnamefont {H.-Y.}\ \bibnamefont {Ko}},
  \bibinfo {author} {\bibfnamefont {A.}~\bibnamefont {Kokalj}}, \bibinfo
  {author} {\bibfnamefont {E.}~\bibnamefont {Küçükbenli}}, \bibinfo {author}
  {\bibfnamefont {M.}~\bibnamefont {Lazzeri}}, \bibinfo {author} {\bibfnamefont
  {M.}~\bibnamefont {Marsili}}, \bibinfo {author} {\bibfnamefont
  {N.}~\bibnamefont {Marzari}}, \bibinfo {author} {\bibfnamefont
  {F.}~\bibnamefont {Mauri}}, \bibinfo {author} {\bibfnamefont {N.~L.}\
  \bibnamefont {Nguyen}}, \bibinfo {author} {\bibfnamefont {H.-V.}\
  \bibnamefont {Nguyen}}, \bibinfo {author} {\bibfnamefont {A.~O.}\
  \bibnamefont {de-la Roza}}, \bibinfo {author} {\bibfnamefont
  {L.}~\bibnamefont {Paulatto}}, \bibinfo {author} {\bibfnamefont
  {S.}~\bibnamefont {Poncé}}, \bibinfo {author} {\bibfnamefont
  {D.}~\bibnamefont {Rocca}}, \bibinfo {author} {\bibfnamefont
  {R.}~\bibnamefont {Sabatini}}, \bibinfo {author} {\bibfnamefont
  {B.}~\bibnamefont {Santra}}, \bibinfo {author} {\bibfnamefont
  {M.}~\bibnamefont {Schlipf}}, \bibinfo {author} {\bibfnamefont {A.~P.}\
  \bibnamefont {Seitsonen}}, \bibinfo {author} {\bibfnamefont {A.}~\bibnamefont
  {Smogunov}}, \bibinfo {author} {\bibfnamefont {I.}~\bibnamefont {Timrov}},
  \bibinfo {author} {\bibfnamefont {T.}~\bibnamefont {Thonhauser}}, \bibinfo
  {author} {\bibfnamefont {P.}~\bibnamefont {Umari}}, \bibinfo {author}
  {\bibfnamefont {N.}~\bibnamefont {Vast}}, \bibinfo {author} {\bibfnamefont
  {X.}~\bibnamefont {Wu}},\ and\ \bibinfo {author} {\bibfnamefont
  {S.}~\bibnamefont {Baroni}},\ }\bibfield  {title} {\bibinfo {title} {Advanced
  capabilities for materials modelling with {QUANTUM ESPRESSO}},\ }\href
  {https://doi.org/10.1088/1361-648X/aa8f79} {\bibfield  {journal} {\bibinfo
  {journal} {Journal of Physics: Condensed Matter}\ }\textbf {\bibinfo {volume}
  {29}},\ \bibinfo {pages} {465901} (\bibinfo {year} {2017})}\BibitemShut
  {NoStop}%
\bibitem [{\citenamefont {Giannozzi}\ \emph {et~al.}(2020)\citenamefont
  {Giannozzi}, \citenamefont {Baseggio}, \citenamefont {Bonfà}, \citenamefont
  {Brunato}, \citenamefont {Car}, \citenamefont {Carnimeo}, \citenamefont
  {Cavazzoni}, \citenamefont {de~Gironcoli}, \citenamefont {Delugas},
  \citenamefont {Ferrari~Ruffino}, \citenamefont {Ferretti}, \citenamefont
  {Marzari}, \citenamefont {Timrov}, \citenamefont {Urru},\ and\ \citenamefont
  {Baroni}}]{giannozzi2020quantum}%
  \BibitemOpen
  \bibfield  {author} {\bibinfo {author} {\bibfnamefont {P.}~\bibnamefont
  {Giannozzi}}, \bibinfo {author} {\bibfnamefont {O.}~\bibnamefont {Baseggio}},
  \bibinfo {author} {\bibfnamefont {P.}~\bibnamefont {Bonfà}}, \bibinfo
  {author} {\bibfnamefont {D.}~\bibnamefont {Brunato}}, \bibinfo {author}
  {\bibfnamefont {R.}~\bibnamefont {Car}}, \bibinfo {author} {\bibfnamefont
  {I.}~\bibnamefont {Carnimeo}}, \bibinfo {author} {\bibfnamefont
  {C.}~\bibnamefont {Cavazzoni}}, \bibinfo {author} {\bibfnamefont
  {S.}~\bibnamefont {de~Gironcoli}}, \bibinfo {author} {\bibfnamefont
  {P.}~\bibnamefont {Delugas}}, \bibinfo {author} {\bibfnamefont
  {F.}~\bibnamefont {Ferrari~Ruffino}}, \bibinfo {author} {\bibfnamefont
  {A.}~\bibnamefont {Ferretti}}, \bibinfo {author} {\bibfnamefont
  {N.}~\bibnamefont {Marzari}}, \bibinfo {author} {\bibfnamefont
  {I.}~\bibnamefont {Timrov}}, \bibinfo {author} {\bibfnamefont
  {A.}~\bibnamefont {Urru}},\ and\ \bibinfo {author} {\bibfnamefont
  {S.}~\bibnamefont {Baroni}},\ }\bibfield  {title} {\bibinfo {title}
  {{{QUANTUM ESPRESSO} toward the exascale}},\ }\href
  {https://doi.org/10.1063/5.0005082} {\bibfield  {journal} {\bibinfo
  {journal} {The Journal of Chemical Physics}\ }\textbf {\bibinfo {volume}
  {152}},\ \bibinfo {pages} {154105} (\bibinfo {year} {2020})}\BibitemShut
  {NoStop}%
\bibitem [{\citenamefont {Perdew}\ \emph {et~al.}(1996)\citenamefont {Perdew},
  \citenamefont {Burke},\ and\ \citenamefont
  {Ernzerhof}}]{perdew1996generalized}%
  \BibitemOpen
  \bibfield  {author} {\bibinfo {author} {\bibfnamefont {J.~P.}\ \bibnamefont
  {Perdew}}, \bibinfo {author} {\bibfnamefont {K.}~\bibnamefont {Burke}},\ and\
  \bibinfo {author} {\bibfnamefont {M.}~\bibnamefont {Ernzerhof}},\ }\bibfield
  {title} {\bibinfo {title} {{G}eneralized {G}radient {A}pproximation {M}ade
  {S}imple},\ }\href {https://doi.org/10.1103/PhysRevLett.77.3865} {\bibfield
  {journal} {\bibinfo  {journal} {Phys. Rev. Lett.}\ }\textbf {\bibinfo
  {volume} {77}},\ \bibinfo {pages} {3865} (\bibinfo {year}
  {1996})}\BibitemShut {NoStop}%
\bibitem [{\citenamefont {{van Setten}}\ \emph {et~al.}(2018)\citenamefont
  {{van Setten}}, \citenamefont {Giantomassi}, \citenamefont {Bousquet},
  \citenamefont {Verstraete}, \citenamefont {Hamann}, \citenamefont {Gonze},\
  and\ \citenamefont {Rignanese}}]{van2018pseudodojo}%
  \BibitemOpen
  \bibfield  {author} {\bibinfo {author} {\bibfnamefont {M.}~\bibnamefont {{van
  Setten}}}, \bibinfo {author} {\bibfnamefont {M.}~\bibnamefont {Giantomassi}},
  \bibinfo {author} {\bibfnamefont {E.}~\bibnamefont {Bousquet}}, \bibinfo
  {author} {\bibfnamefont {M.}~\bibnamefont {Verstraete}}, \bibinfo {author}
  {\bibfnamefont {D.}~\bibnamefont {Hamann}}, \bibinfo {author} {\bibfnamefont
  {X.}~\bibnamefont {Gonze}},\ and\ \bibinfo {author} {\bibfnamefont {G.-M.}\
  \bibnamefont {Rignanese}},\ }\bibfield  {title} {\bibinfo {title} {The
  {PSEUDODOJO}: {T}raining and grading a 85 element optimized norm-conserving
  pseudopotential table},\ }\href
  {https://doi.org/https://doi.org/10.1016/j.cpc.2018.01.012} {\bibfield
  {journal} {\bibinfo  {journal} {Computer Physics Communications}\ }\textbf
  {\bibinfo {volume} {226}},\ \bibinfo {pages} {39} (\bibinfo {year}
  {2018})}\BibitemShut {NoStop}%
\bibitem [{\citenamefont {Yun}\ \emph {et~al.}(2012)\citenamefont {Yun},
  \citenamefont {Han}, \citenamefont {Hong}, \citenamefont {Kim},\ and\
  \citenamefont {Lee}}]{yun2012thickness}%
  \BibitemOpen
  \bibfield  {author} {\bibinfo {author} {\bibfnamefont {W.~S.}\ \bibnamefont
  {Yun}}, \bibinfo {author} {\bibfnamefont {S.}~\bibnamefont {Han}}, \bibinfo
  {author} {\bibfnamefont {S.~C.}\ \bibnamefont {Hong}}, \bibinfo {author}
  {\bibfnamefont {I.~G.}\ \bibnamefont {Kim}},\ and\ \bibinfo {author}
  {\bibfnamefont {J.}~\bibnamefont {Lee}},\ }\bibfield  {title} {\bibinfo
  {title} {{Thickness and strain effects on electronic structures of transition
  metal dichalcogenides: 2H-MX${}_2$ semiconductors (M=Mo, W; X=S, Se, Te)}},\
  }\href@noop {} {\bibfield  {journal} {\bibinfo  {journal} {Physical Review
  B—Condensed Matter and Materials Physics}\ }\textbf {\bibinfo {volume}
  {85}},\ \bibinfo {pages} {033305} (\bibinfo {year} {2012})}\BibitemShut
  {NoStop}%
\bibitem [{\citenamefont {Monkhorst}\ and\ \citenamefont
  {Pack}(1976)}]{monkhorst1976special}%
  \BibitemOpen
  \bibfield  {author} {\bibinfo {author} {\bibfnamefont {H.~J.}\ \bibnamefont
  {Monkhorst}}\ and\ \bibinfo {author} {\bibfnamefont {J.~D.}\ \bibnamefont
  {Pack}},\ }\bibfield  {title} {\bibinfo {title} {{S}pecial points for
  {B}rillouin-zone integrations},\ }\href
  {https://doi.org/10.1103/PhysRevB.13.5188} {\bibfield  {journal} {\bibinfo
  {journal} {Phys. Rev. B}\ }\textbf {\bibinfo {volume} {13}},\ \bibinfo
  {pages} {5188} (\bibinfo {year} {1976})}\BibitemShut {NoStop}%
\bibitem [{\citenamefont {Reining}(2018)}]{reining2018gw}%
  \BibitemOpen
  \bibfield  {author} {\bibinfo {author} {\bibfnamefont {L.}~\bibnamefont
  {Reining}},\ }\bibfield  {title} {\bibinfo {title} {The {GW} approximation:
  content, successes and limitations},\ }\href
  {https://doi.org/https://doi.org/10.1002/wcms.1344} {\bibfield  {journal}
  {\bibinfo  {journal} {WIREs Comput. Mol. Sci.}\ }\textbf {\bibinfo {volume}
  {8}},\ \bibinfo {pages} {e1344} (\bibinfo {year} {2018})}\BibitemShut
  {NoStop}%
\bibitem [{\citenamefont {Onida}\ \emph {et~al.}(2002)\citenamefont {Onida},
  \citenamefont {Reining},\ and\ \citenamefont {Rubio}}]{onida2002electronic}%
  \BibitemOpen
  \bibfield  {author} {\bibinfo {author} {\bibfnamefont {G.}~\bibnamefont
  {Onida}}, \bibinfo {author} {\bibfnamefont {L.}~\bibnamefont {Reining}},\
  and\ \bibinfo {author} {\bibfnamefont {A.}~\bibnamefont {Rubio}},\ }\bibfield
   {title} {\bibinfo {title} {{E}lectronic excitations: density-functional
  versus many-body {G}reen’s-function approaches},\ }\href
  {https://doi.org/10.1103/RevModPhys.74.601} {\bibfield  {journal} {\bibinfo
  {journal} {Rev. Mod. Phys.}\ }\textbf {\bibinfo {volume} {74}},\ \bibinfo
  {pages} {601} (\bibinfo {year} {2002})}\BibitemShut {NoStop}%
\bibitem [{\citenamefont {Strinati}(1988)}]{strinati1988application}%
  \BibitemOpen
  \bibfield  {author} {\bibinfo {author} {\bibfnamefont {G.}~\bibnamefont
  {Strinati}},\ }\bibfield  {title} {\bibinfo {title} {{A}pplication of the
  {G}reen’s functions method to the study of the optical properties of
  semiconductors},\ }\href {https://doi.org/10.1007/BF02725962} {\bibfield
  {journal} {\bibinfo  {journal} {La Rivista del Nuovo Cimento (1978-1999)}\
  }\textbf {\bibinfo {volume} {11}},\ \bibinfo {pages} {1} (\bibinfo {year}
  {1988})}\BibitemShut {NoStop}%
\bibitem [{\citenamefont {Marini}\ \emph {et~al.}(2009)\citenamefont {Marini},
  \citenamefont {Hogan}, \citenamefont {Grüning},\ and\ \citenamefont
  {Varsano}}]{marini_yambo_2009}%
  \BibitemOpen
  \bibfield  {author} {\bibinfo {author} {\bibfnamefont {A.}~\bibnamefont
  {Marini}}, \bibinfo {author} {\bibfnamefont {C.}~\bibnamefont {Hogan}},
  \bibinfo {author} {\bibfnamefont {M.}~\bibnamefont {Grüning}},\ and\
  \bibinfo {author} {\bibfnamefont {D.}~\bibnamefont {Varsano}},\ }\bibfield
  {title} {\bibinfo {title} {{{y}ambo: {A}n ab initio tool for excited state
  calculations}},\ }\href
  {https://doi.org/https://doi.org/10.1016/j.cpc.2009.02.003} {\bibfield
  {journal} {\bibinfo  {journal} {Computer Physics Communications}\ }\textbf
  {\bibinfo {volume} {180}},\ \bibinfo {pages} {1392} (\bibinfo {year}
  {2009})}\BibitemShut {NoStop}%
\bibitem [{\citenamefont {Sangalli}\ \emph {et~al.}(2019)\citenamefont
  {Sangalli}, \citenamefont {Ferretti}, \citenamefont {Miranda}, \citenamefont
  {Attaccalite}, \citenamefont {Marri}, \citenamefont {Cannuccia},
  \citenamefont {Melo}, \citenamefont {Marsili}, \citenamefont {Paleari},
  \citenamefont {Marrazzo}, \citenamefont {Prandini}, \citenamefont {Bonfà},
  \citenamefont {Atambo}, \citenamefont {Affinito}, \citenamefont {Palummo},
  \citenamefont {Molina-Sánchez}, \citenamefont {Hogan}, \citenamefont
  {Grüning}, \citenamefont {Varsano},\ and\ \citenamefont
  {Marini}}]{sangalli2019many}%
  \BibitemOpen
  \bibfield  {author} {\bibinfo {author} {\bibfnamefont {D.}~\bibnamefont
  {Sangalli}}, \bibinfo {author} {\bibfnamefont {A.}~\bibnamefont {Ferretti}},
  \bibinfo {author} {\bibfnamefont {H.}~\bibnamefont {Miranda}}, \bibinfo
  {author} {\bibfnamefont {C.}~\bibnamefont {Attaccalite}}, \bibinfo {author}
  {\bibfnamefont {I.}~\bibnamefont {Marri}}, \bibinfo {author} {\bibfnamefont
  {E.}~\bibnamefont {Cannuccia}}, \bibinfo {author} {\bibfnamefont
  {P.}~\bibnamefont {Melo}}, \bibinfo {author} {\bibfnamefont {M.}~\bibnamefont
  {Marsili}}, \bibinfo {author} {\bibfnamefont {F.}~\bibnamefont {Paleari}},
  \bibinfo {author} {\bibfnamefont {A.}~\bibnamefont {Marrazzo}}, \bibinfo
  {author} {\bibfnamefont {G.}~\bibnamefont {Prandini}}, \bibinfo {author}
  {\bibfnamefont {P.}~\bibnamefont {Bonfà}}, \bibinfo {author} {\bibfnamefont
  {M.~O.}\ \bibnamefont {Atambo}}, \bibinfo {author} {\bibfnamefont
  {F.}~\bibnamefont {Affinito}}, \bibinfo {author} {\bibfnamefont
  {M.}~\bibnamefont {Palummo}}, \bibinfo {author} {\bibfnamefont
  {A.}~\bibnamefont {Molina-Sánchez}}, \bibinfo {author} {\bibfnamefont
  {C.}~\bibnamefont {Hogan}}, \bibinfo {author} {\bibfnamefont
  {M.}~\bibnamefont {Grüning}}, \bibinfo {author} {\bibfnamefont
  {D.}~\bibnamefont {Varsano}},\ and\ \bibinfo {author} {\bibfnamefont
  {A.}~\bibnamefont {Marini}},\ }\bibfield  {title} {\bibinfo {title}
  {{M}any-body perturbation theory calculations using the yambo code},\ }\href
  {https://doi.org/10.1088/1361-648X/ab15d0} {\bibfield  {journal} {\bibinfo
  {journal} {Journal of Physics: Condensed Matter}\ }\textbf {\bibinfo {volume}
  {31}},\ \bibinfo {pages} {325902} (\bibinfo {year} {2019})}\BibitemShut
  {NoStop}%
\bibitem [{\citenamefont {Marsili}\ \emph {et~al.}(2021)\citenamefont
  {Marsili}, \citenamefont {Molina-S\'anchez}, \citenamefont {Palummo},
  \citenamefont {Sangalli},\ and\ \citenamefont
  {Marini}}]{marsili2021spinorial}%
  \BibitemOpen
  \bibfield  {author} {\bibinfo {author} {\bibfnamefont {M.}~\bibnamefont
  {Marsili}}, \bibinfo {author} {\bibfnamefont {A.}~\bibnamefont
  {Molina-S\'anchez}}, \bibinfo {author} {\bibfnamefont {M.}~\bibnamefont
  {Palummo}}, \bibinfo {author} {\bibfnamefont {D.}~\bibnamefont {Sangalli}},\
  and\ \bibinfo {author} {\bibfnamefont {A.}~\bibnamefont {Marini}},\
  }\bibfield  {title} {\bibinfo {title} {{S}pinorial formulation of the
  ${GW}$-{BSE} equations and spin properties of excitons in two-dimensional
  transition metal dichalcogenides},\ }\href
  {https://doi.org/10.1103/PhysRevB.103.155152} {\bibfield  {journal} {\bibinfo
   {journal} {Phys. Rev. B}\ }\textbf {\bibinfo {volume} {103}},\ \bibinfo
  {pages} {155152} (\bibinfo {year} {2021})}\BibitemShut {NoStop}%
\bibitem [{\citenamefont {Stan}\ \emph {et~al.}(2009)\citenamefont {Stan},
  \citenamefont {Dahlen},\ and\ \citenamefont {Van~Leeuwen}}]{stan2009levels}%
  \BibitemOpen
  \bibfield  {author} {\bibinfo {author} {\bibfnamefont {A.}~\bibnamefont
  {Stan}}, \bibinfo {author} {\bibfnamefont {N.~E.}\ \bibnamefont {Dahlen}},\
  and\ \bibinfo {author} {\bibfnamefont {R.}~\bibnamefont {Van~Leeuwen}},\
  }\bibfield  {title} {\bibinfo {title} {{Levels of self-consistency in the GW
  approximation}},\ }\href@noop {} {\bibfield  {journal} {\bibinfo  {journal}
  {The Journal of chemical physics}\ }\textbf {\bibinfo {volume} {130}}
  (\bibinfo {year} {2009})}\BibitemShut {NoStop}%
\bibitem [{\citenamefont {Rojas}\ \emph {et~al.}(1995)\citenamefont {Rojas},
  \citenamefont {Godby},\ and\ \citenamefont {Needs}}]{PPA_yambo}%
  \BibitemOpen
  \bibfield  {author} {\bibinfo {author} {\bibfnamefont {H.~N.}\ \bibnamefont
  {Rojas}}, \bibinfo {author} {\bibfnamefont {R.~W.}\ \bibnamefont {Godby}},\
  and\ \bibinfo {author} {\bibfnamefont {R.~J.}\ \bibnamefont {Needs}},\
  }\bibfield  {title} {\bibinfo {title} {Space-time method for ab initio
  calculations of self-energies and dielectric response functions of solids},\
  }\href {https://doi.org/10.1103/PhysRevLett.74.1827} {\bibfield  {journal}
  {\bibinfo  {journal} {Phys. Rev. Lett.}\ }\textbf {\bibinfo {volume} {74}},\
  \bibinfo {pages} {1827} (\bibinfo {year} {1995})}\BibitemShut {NoStop}%
\bibitem [{\citenamefont {Bohm}\ and\ \citenamefont {Pines}(1953)}]{RPA}%
  \BibitemOpen
  \bibfield  {author} {\bibinfo {author} {\bibfnamefont {D.}~\bibnamefont
  {Bohm}}\ and\ \bibinfo {author} {\bibfnamefont {D.}~\bibnamefont {Pines}},\
  }\bibfield  {title} {\bibinfo {title} {{A Collective Description of Electron
  Interactions: III. Coulomb Interactions in a Degenerate Electron Gas}},\
  }\href {https://doi.org/10.1103/PhysRev.92.609} {\bibfield  {journal}
  {\bibinfo  {journal} {Phys. Rev.}\ }\textbf {\bibinfo {volume} {92}},\
  \bibinfo {pages} {609} (\bibinfo {year} {1953})}\BibitemShut {NoStop}%
\bibitem [{\citenamefont {Guandalini}\ \emph {et~al.}(2023)\citenamefont
  {Guandalini}, \citenamefont {D’Amico}, \citenamefont {Ferretti},\ and\
  \citenamefont {Varsano}}]{guandalini2022efficient}%
  \BibitemOpen
  \bibfield  {author} {\bibinfo {author} {\bibfnamefont {A.}~\bibnamefont
  {Guandalini}}, \bibinfo {author} {\bibfnamefont {P.}~\bibnamefont
  {D’Amico}}, \bibinfo {author} {\bibfnamefont {A.}~\bibnamefont
  {Ferretti}},\ and\ \bibinfo {author} {\bibfnamefont {D.}~\bibnamefont
  {Varsano}},\ }\bibfield  {title} {\bibinfo {title} {Efficient {GW}
  calculations in two dimensional materials through a stochastic integration of
  the screened potential},\ }\href {https://doi.org/10.1038/s41524-023-00989-7}
  {\bibfield  {journal} {\bibinfo  {journal} {npj Computational Materials}\
  }\textbf {\bibinfo {volume} {9}},\ \bibinfo {pages} {44} (\bibinfo {year}
  {2023})}\BibitemShut {NoStop}%
\bibitem [{\citenamefont {Bruneval}\ and\ \citenamefont
  {Gonze}(2008)}]{bruneval2008accurate}%
  \BibitemOpen
  \bibfield  {author} {\bibinfo {author} {\bibfnamefont {F.}~\bibnamefont
  {Bruneval}}\ and\ \bibinfo {author} {\bibfnamefont {X.}~\bibnamefont
  {Gonze}},\ }\bibfield  {title} {\bibinfo {title} {{A}ccurate ${GW}$
  self-energies in a plane-wave basis using only a few empty states: {T}owards
  large systems},\ }\href {https://doi.org/10.1103/PhysRevB.78.085125}
  {\bibfield  {journal} {\bibinfo  {journal} {Phys. Rev. B}\ }\textbf {\bibinfo
  {volume} {78}},\ \bibinfo {pages} {085125} (\bibinfo {year}
  {2008})}\BibitemShut {NoStop}%
\end{thebibliography}
\end{document}